\documentclass[aps,prd,twoside,onecolumn,superscriptaddress,floatfix,nofootinbib]{revtex4}
\usepackage{amsmath}
\usepackage{bm}
\usepackage{xcolor}

\newcommand\beq{\begin{equation}}
\newcommand\eeq{\end{equation}}
\newcommand\beqn{\begin{eqnarray}}
\newcommand\eeqn{\end{eqnarray}}

\newcommand{\ba}{\begin{eqnarray}}
\newcommand{\ea}{\end{eqnarray}}
\newcommand{\be}{\begin{equation}}
\newcommand{\ee}{\end{equation}}

\newcommand\lsim{\mathrel{\rlap{\lower4pt\hbox{\hskip1pt$\sim$}}
        \raise1pt\hbox{$<$}}}
\newcommand\gsim{\mathrel{\rlap{\lower4pt\hbox{\hskip1pt$\sim$}}
        \raise1pt\hbox{$>$}}}

\renewcommand{\vec}[1]{\boldsymbol{\mathbf{#1}}}

\newcommand{\jcap}{{J.~Cosm.~Astrop.~Phys.}}

\newcommand{\aap}{{Astron.~Astrophys.}}
\newcommand{\apjl}{{Astrophys.~J.~Lett.}}
\newcommand{\apjs}{{Astrophys.~J.~Supp.}}

\newcommand{\mnras}{{Mon.~Not.~R.~Astron.~Soc.}}
\newcommand{\njp}{{New~J.~Phys.}}
\usepackage{graphicx}
\usepackage[large]{subfigure}
\usepackage{amssymb, amsmath}
\usepackage[amssymb]{SIunits}
\usepackage{epstopdf}
\usepackage{hyperref}
\usepackage{aas_macros}
\usepackage{natbib}

\begin{document}

\title{The Atacama Cosmology Telescope: A Measurement of the Thermal Sunyaev-Zel'dovich One-Point PDF}
 \author{J.~Colin Hill}
  \affiliation{Dept.~of Astronomy, Pupin Hall, Columbia University, New York, NY, USA 10027}
 \affiliation{Dept.~of Astrophysical Sciences, Peyton Hall, Princeton University, Princeton, NJ, USA 08544}
 \author{Blake D.~Sherwin}
 \affiliation{Dept.~of Physics, University of California, Berkeley, CA, USA 94720}
 \affiliation{Miller Institute for Basic Research in Science, University of California, Berkeley, CA, USA 94720}
 \affiliation{Dept.~of Physics, Princeton University, Princeton, NJ, USA 08544}
 \author{Kendrick~M.~Smith}
 \affiliation{Perimeter Institute for Theoretical Physics, Waterloo ON N2L 2Y5, Canada}
 \affiliation{Dept.~of Astrophysical Sciences, Peyton Hall, Princeton University, Princeton, NJ, USA 08544}
\author{Graeme~E.~Addison}\affiliation{Dept.~of Physics and Astronomy, University of British Columbia, Vancouver, BC, Canada V6T 1Z4}
\author{Nick~Battaglia}
\affiliation{McWilliams Center for Cosmology, Dept.~of Physics, Carnegie Mellon University, Pittsburgh, PA, USA 15213}
\affiliation{Dept.~of Astrophysical Sciences, Peyton Hall, Princeton University, Princeton, NJ, USA 08544}
\author{Elia~S.~Battistelli}
\affiliation{Dept.~of Physics, ``Sapienza'' University of Rome, Piazzale Aldo Moro 5, I-00185 Rome, Italy}
\affiliation{Dept.~of Physics and Astronomy, University of British Columbia, Vancouver, BC, Canada V6T 1Z4}
\author{J.~Richard~Bond}\affiliation{CITA, University of Toronto, Toronto, ON, Canada M5S 3H8}
\author{Erminia~Calabrese}\affiliation{Sub-dept.~of Astrophysics, University of Oxford, Keble Road, Oxford OX1 3RH, UK}
\author{Mark~J.~Devlin}\affiliation{Dept.~of Physics and Astronomy, University of Pennsylvania, Philadelphia, PA, USA 19104}
\author{Joanna~Dunkley}\affiliation{Sub-dept.~of Astrophysics, University of Oxford, Keble Road, Oxford OX1 3RH, UK}
\author{Rolando~D\"{u}nner}
\affiliation{Departamento de Astronom{\'{i}}a y Astrof{\'{i}}sica, Pontific\'{i}a Univ. Cat\'{o}lica, Casilla 306, Santiago 22, Chile}
\author{Thomas~Essinger-Hileman}
\affiliation{Dept.~of Physics and Astronomy, The Johns Hopkins University, Baltimore, MD, USA 21218}
\author{Megan~B.~Gralla}
\affiliation{Dept.~of Physics and Astronomy, The Johns Hopkins University, Baltimore, MD, USA 21218}
\affiliation{Harvard-Smithsonian Center for Astrophysics, 60 Garden Street, Cambridge, MA, USA 02138}
\author{Amir~Hajian}\affiliation{CITA, University of Toronto, Toronto, ON, Canada M5S 3H8}
\author{Matthew~Hasselfield}
 \affiliation{Dept.~of Astrophysical Sciences, Peyton Hall, Princeton University, Princeton, NJ, USA 08544}
\affiliation{Dept.~of Physics and Astronomy, University of British Columbia, Vancouver, BC, Canada V6T 1Z4}
\author{Adam~D.~Hincks}\affiliation{Dept.~of Physics and Astronomy, University of British Columbia, Vancouver, BC, Canada V6T 1Z4}
\author{Ren\'ee~Hlo\^{z}ek}
 \affiliation{Dept.~of Astrophysical Sciences, Peyton Hall, Princeton University, Princeton, NJ, USA 08544}
\author{John~P.~Hughes}\affiliation{Dept.~of Physics and Astronomy, Rutgers, The State University of New Jersey, Piscataway, NJ USA 08854}
\author{Arthur~Kosowsky}\affiliation{Dept.~of Physics and Astronomy, University of Pittsburgh, Pittsburgh, PA, USA 15260}
\author{Thibaut~Louis}\affiliation{Sub-dept.~of Astrophysics, University of Oxford, Keble Road, Oxford OX1 3RH, UK}
\author{Danica~Marsden}\affiliation{Dept.~of Physics and Astronomy, University of Pennsylvania, Philadelphia, PA, USA 19104}
\author{Kavilan~Moodley}\affiliation{Astrophysics and Cosmology Research Unit, University of KwaZulu-Natal, Durban, 4041, South Africa}
\author{Michael~D.~Niemack}
\affiliation{Dept.~of Physics, Cornell University, Ithaca, NY, USA 14853}
\author{Lyman~A.~Page}\affiliation{Dept.~of Physics, Princeton University, Princeton, NJ, USA 08544}
\author{Bruce~Partridge}\affiliation{Dept.~of Physics and Astronomy, Haverford College, Haverford, PA, USA 19041}
\author{Benjamin~Schmitt}\affiliation{Dept.~of Physics and Astronomy, University of Pennsylvania, Philadelphia, PA, USA 19104}
\author{Neelima~Sehgal}\affiliation{Physics and Astronomy Dept., Stony Brook University, Stony Brook, NY, USA 11794}
\author{Jonathan~L.~Sievers}
\affiliation{Astrophysics and Cosmology Research Unit, University of KwaZulu-Natal, Durban, 4041, South Africa}
\affiliation{Dept.~of Physics, Princeton University, Princeton, NJ, USA 08544}
\author{David~N.~Spergel}
\affiliation{Dept.~of Astrophysical Sciences, Peyton Hall, Princeton University, Princeton, NJ, USA 08544}
\author{Suzanne~T.~Staggs}\affiliation{Dept.~of Physics, Princeton University, Princeton, NJ, USA 08544}
\author{Daniel~S.~Swetz}
\affiliation{Dept.~of Physics and Astronomy, University of Pennsylvania, Philadelphia, PA, USA 19104}
\affiliation{NIST Quantum Devices Group, 325 Broadway Mailcode 817.03, Boulder, CO, USA 80305}
\author{Robert Thornton}
\affiliation{Dept.~of Physics and Astronomy, University of Pennsylvania, Philadelphia, PA, USA 19104}
\affiliation{Dept.~of Physics, West Chester University of Pennsylvania, West Chester, PA, USA 19383}
\author{Hy~Trac}
\affiliation{McWilliams Center for Cosmology, Dept.~of Physics, Carnegie Mellon University, Pittsburgh, PA, USA 15213}
\author{Edward~J.~Wollack}\affiliation{NASA/Goddard Space Flight Center, Greenbelt, MD, USA 20771}

\begin{abstract}
We present a measurement of the one-point probability distribution
function (PDF) of the thermal Sunyaev-Zel'dovich (tSZ) decrement in
the pixel temperature histogram of filtered $148$ GHz sky maps from the
Atacama Cosmology Telescope (ACT). The PDF includes the signal from
all galaxy clusters in the map, including objects below the
signal-to-noise threshold for individual detection, making it a
particularly sensitive probe of the amplitude of matter density
perturbations, $\sigma_8$. We use a combination of analytic halo model calculations and
numerical simulations to compute the theoretical tSZ PDF and its
covariance matrix, accounting for all noise sources and including
relativistic corrections. From the measured ACT $148$ GHz PDF alone,
we find $\sigma_8 = 0.793 \pm 0.018$, with additional systematic errors of $\pm 0.017$ due to uncertainty in 
intracluster medium gas physics and $\pm 0.006$ due to uncertainty in infrared
point source contamination. Using effectively the same data set, the statistical error here is a factor of two lower than that found
in ACT's previous $\sigma_8$ determination based solely on the skewness
of the tSZ signal. In future temperature maps with higher sensitivity,
the tSZ PDF will break the degeneracy between intracluster medium gas
physics and cosmological parameters.
\end{abstract}
\maketitle

\section{Introduction}
In recent years, studies of the cosmic microwave background (CMB) temperature anisotropies have progressed beyond measurements of the primordial fluctuations seeded by inhomogeneities in the baryon-photon plasma at $z \approx 1100$, moving on to the study of secondary fluctuations induced by various physical processes at $z \lsim 10$.  This progress has been possible due to significant improvements in resolution and sensitivity, as demonstrated in the current generation of CMB experiments, including, for example, the Atacama Cosmology Telescope (ACT)/ACTPol~\cite{Swe11,Niemacketal2010}, South Pole Telescope (SPT)/SPTPol~\cite{Car11,Austermannetal2012}, \emph{Planck}~\cite{Planck2013overview}, and {\sc Polarbear}~\cite{Kermishetal2012}.  Of particular note is the rapid growth in measurements of the Sunyaev-Zel'dovich (SZ) effect.  The SZ effect is a spectral distortion induced in the CMB by the scattering of CMB photons off free electrons~\cite{Zel69,Sunyaev-Zeldovich1970}.  This encompasses the kinetic SZ (kSZ) effect (first detected recently in ACT data~\cite{Handetal2012}), which results from the scattering of CMB photons off electrons possessing a non-zero peculiar velocity along the line-of-sight, and the thermal SZ (tSZ) effect, which results from the scattering of CMB photons off hot electrons, thereby altering the CMB spectrum.  The electrons sourcing the tSZ effect are predominantly found in the intracluster medium (ICM) of massive galaxy clusters.  The tSZ effect has now been measured over a wide range of halo masses and redshifts~(e.g.,~\cite{Hasselfieldetal2013,Grallaetal2014,Reichardtetal2013,Planck2013counts,Planck2013LBG}), both in direct observations and blind surveys, and has also been studied indirectly through its contribution to the power spectrum and higher-point functions of CMB temperature maps~\cite{Sieversetal2013,Storyetal2013,Wilsonetal2012,Crawfordetal2014,Planck2013ymap}.

This paper is focused in particular on an indirect, statistical approach to measuring the tSZ signal by using the one-point probability distribution function (PDF) of the tSZ-induced temperature fluctuations.  In practice, this corresponds to a measurement of the histogram of pixel values in a CMB temperature map.  In earlier work, ACT data were used to measure the tSZ signal through the skewness of this histogram~\cite{Wilsonetal2012} (hereafter W12); here, we work directly with the tSZ PDF, rather than moments of the tSZ field.  We measure the tSZ PDF in Wiener-filtered ACT 148 GHz data in a $\approx 324$ square degree region along the celestial equator.  We interpret the measurement using a combination of analytic halo model calculations and numerical simulations, and use the results to constrain cosmological parameters, with appropriate marginalization of astrophysical uncertainties.  We consider systematics from dust and other non-tSZ foregrounds, finding minimal effects with the exception of contamination from IR sources ``filling in'' tSZ decrements at 148 GHz.  The overall goal of the PDF analysis is to unify and optimize existing approaches to tSZ statistics in a conceptually straightforward way.

Indirect observables such as the tSZ PDF, tSZ power spectrum, or tSZ bispectrum possess some advantages over direct methods (i.e., finding and counting clusters).  For example, no selection function is applied to the map to find clusters, and hence no selection effect-related systematics arise, such as Malmquist or Eddington bias.  In addition, no choice of ``aperture'' is required within which a cluster mass is defined (though this can be a useful intermediate step in halo model calculations of tSZ statistics), and indeed no measurement of individual cluster masses is performed.  
Likewise, the use of ``average'' gas pressure profile prescriptions is sensible when computing tSZ statistics, since one is computing a population-level quantity (i.e., there is no need to apply an average/universal profile to potentially anomalous individual objects).  Lastly, as has been known for many years, tSZ statistics are very sensitive to $\sigma_8$, the rms amplitude of matter density fluctuations on scales of $8$ Mpc$/h$~(e.g.,~\cite{Komatsu-Seljak2002,Hill-Sherwin2013,Hill-Pajer2013,Bhattacharyaetal2012}).  The tSZ one-point PDF inherits all of these advantages, although it comes with the same disadvantages as other indirect tSZ probes --- for example, no redshift information is available, as the entire line-of-sight is integrated over, and no follow-up observations of individual systems are made.  In addition, like other indirect methods, the tSZ PDF possesses the simultaneous advantage and disadvantage of sensitivity to lower-mass, higher-redshift halos than those found in cluster counts.  This is an advantage because of the raw increase in statistical power from this sensitivity; it is a disadvantage because the ICM gas physics in these systems is more uncertain than that in massive, low-redshift systems with deep X-ray, optical, and lensing observations~(e.g.,~\cite{Shawetal2010,Battagliaetal2012,Hill-Sherwin2013,Bhattacharyaetal2012}).  However, in a more optimistic sense, the tSZ PDF and other tSZ statistics allow for the opportunity to constrain the gas physics in these systems, which have not been observed in targeted studies.

Nearly all cosmological constraints derived from measurements of tSZ statistics thus far are based solely on the one-halo term, because it dominates the total signal over a wide range of angular scales.  In effect, these measurements are just indirect ways of counting clusters, including those below the signal-to-noise ratio (SNR) threshold for direct detection.  Spatial clustering information in the two-, three-, and higher-point terms is not measured~\cite{Komatsu-Kitayama1999,Hill-Pajer2013}.  The only exceptions are recent hints of the two-halo term in the \emph{Planck} Compton-$y$ map power spectrum~\cite{Planck2013ymap} and measurements of the two-halo term in the cross-correlation of tSZ and gravitational lensing maps~\cite{Hill-Spergel2014,vanWaerbekeetal2014}.  This situation thus motivates the consideration of a statistic that fully uses the information in the tSZ one-halo term --- the one-point PDF.

The tSZ PDF includes the information contained in all zero-lag moments of the tSZ field.  It has been shown that higher-point tSZ statistics scale with increasingly high powers of $\sigma_8$, with $\langle T^2 \rangle \propto \sigma_8^{7-8}$, $\langle T^3 \rangle \propto \sigma_8^{10-12}$, $\langle T^4 \rangle \propto \sigma_8^{13-15}$, and continuing on in this manner~\cite{Hill-Sherwin2013,Bhattacharyaetal2012}.  Here and throughout this paper, $T(\hat{\mathbf{n}})$ refers to the temperature fluctuation in direction $\hat{\mathbf{n}}$ on the sky, $T(\hat{\mathbf{n}}) \equiv \mathcal{T}(\hat{\mathbf{n}}) - \mathcal{T}_{\mathrm{CMB}}$, where $\mathcal{T}(\hat{\mathbf{n}})$ is temperature and $\mathcal{T}_{\mathrm{CMB}} = 2.726$ K is the CMB temperature today.  The underlying reason for this sensitivity to $\sigma_8$ is that increasingly higher-point tSZ statistics are dominated by contributions from increasingly rare, more massive clusters that lie deeper in the exponential tail of the mass function.  The tSZ PDF includes information from all of these higher moments, which eventually reach a sample variance-dominated limit as one proceeds to consider progressively higher-point statistics.  Apart from discarding redshift information, the tSZ PDF is thus likely close to an optimal one-halo-term tSZ statistic for constraining cosmological parameters.

Moreover, by naturally including information from all moments of the tSZ field, the tSZ PDF possesses the ability to break the degeneracy usually present in tSZ statistics between variations in cosmological parameters and variations in the ICM gas physics model.  This idea is a natural generalization of the method presented in~\cite{Hill-Sherwin2013} --- the moments comprising the PDF depend differently on the cosmological and ICM parameters, allowing degeneracies between these parameters to be broken.  
In this analysis, the data are not quite at the level needed to strongly break the cosmology--ICM degeneracy.  The problem is made more challenging by the highly correlated, non-Gaussian nature of the PDF likelihood function (see Section~\ref{sec:interpretation} below), which we simplify by combining many of the bins in the tail of the tSZ PDF.  With a more sophisticated approach to the likelihood function and wider, deeper maps, future measurements of the tSZ PDF should allow for a stronger breaking of the cosmology--ICM degeneracy.  

Theoretically, the tSZ PDF requires essentially the same modeling inputs as other tSZ statistics, in particular an understanding of the ICM gas pressure profile over a fairly wide range of masses and redshifts.  For the ACT Equatorial map sensitivity and resolution~\cite{Dasetal2014}, the PDF is mostly dominated by systems at preferentially lower redshifts and higher masses than those that dominate the tSZ power spectrum at $\ell = 3000$~(e.g.,~\cite{Tracetal2011,Battagliaetal2012}), implying that the modeling uncertainties should not be overwhelming, similar to the situation with the tSZ skewness and kurtosis~\cite{Hill-Sherwin2013,Bhattacharyaetal2012}.  In addition, we largely circumvent non-tSZ contamination by working with only the $T < 0$ side of the 148 GHz ACT PDF, in which the non-Gaussian contributions are dominated by the tSZ signal (recall that the tSZ effect yields a decrement in the observed CMB temperature at this frequency).  However, a potential difficulty arises from the currently poorly constrained correlation between the cosmic infrared background (CIB) and the tSZ field~\cite{Addisonetal2012,Mesingeretal2012}.  The CIB is comprised of the cumulative infrared (IR) emission from billions of unresolved dusty star-forming galaxies, some of which may be spatially co-located in or near the galaxy groups and clusters sourcing the tSZ signal.  Fortunately, since the tSZ PDF is a one-point statistic, it is unaffected by the clustering (two-halo) term in the CIB--tSZ correlation; the only possible issue comes from IR sources ``filling in'' the CMB temperature decrements induced by the tSZ effect at 148 GHz.  We investigate this issue using the ACT 218 GHz maps in Section~\ref{sec:interpretation}.

To our knowledge, this paper represents the first work to use the tSZ PDF as a cosmological probe.  Similar approaches have been investigated in the weak lensing and large-scale structure literature~(e.g.,~\cite{Jain-vanWaerbeke2000,Neyrinck2014}), and our method is similar to the measurement of ``peak counts'' or the convergence PDF in weak lensing maps~(e.g.~\cite{Kratochviletal2010}), as well as the traditional $P(D)$ analysis used in radio point source studies for decades~\cite{Scheuer1957,Condon1974}.  The first theoretical calculation of the tSZ PDF was performed in~\cite{Bond-Myers1996} using the ``peak-patch'' picture of cosmic structure formation, a somewhat different approach than we adopt here (see Section~\ref{sec:theory}).  In subsequent years, a number of simulation groups measured the PDF of the tSZ field in cosmological hydrodynamics simulations~\cite{daSilvaetal2000,Seljaketal2001,Springeletal2001,Springeletal2001ERR,Refregier-Teyssier2002,Zhangetal2002}.  These groups noted the significant non-Gaussianity in the tSZ PDF and suggested it as a way to distinguish the tSZ fluctuations from the (Gaussian) primary CMB field in a single-frequency experiment.  Other analyses at the time attempted to compute the tSZ PDF analytically using an Edgeworth expansion in combination with halo model calculations of tSZ moments~\cite{Cooray2000}; unfortunately, this expansion is numerically unstable when applied to the tSZ PDF (with terms beyond the skewness) due to the oscillating sign of each term and the large magnitude of the higher tSZ moments~\cite{Hill-Sherwin2013}.  The Edgeworth expansion likely does not converge when applied to the tSZ PDF.  Thus, instead of working in terms of an expansion around a Gaussian distribution or using perturbation theory, in this paper  we introduce a method to compute the tSZ PDF directly using the halo model, similar to approaches already used to calculate the tSZ power spectrum~\cite{Komatsu-Seljak2002}, tSZ bispectrum~\cite{Bhattacharyaetal2012}, and real-space tSZ moments~\cite{Hill-Sherwin2013}.  We then use these calculations to interpret the tSZ PDF measured with ACT data, yielding constraints on cosmological parameters.  The data set used here is nearly identical to that used in the ACT tSZ skewness analysis of~W12; the improved precision of our results compared to that work is due to the use of a more powerful tSZ statistic.

The rest of this paper is organized as follows.  Section~\ref{sec:data} describes the ACT 148 GHz data used in the analysis and presents its pixel temperature histogram.  Section~\ref{sec:theory} describes our halo model-based theoretical approach to the tSZ PDF and verifies its accuracy using numerical simulations.  We also describe how to account for noise in the PDF and outline our models for the various non-tSZ contributions to the ACT PDF.  In addition, we investigate the sensitivity of the tSZ PDF to cosmological and astrophysical parameters, and calculate the contributions to the signal from different mass and redshift ranges.  Section~\ref{sec:sims} presents the simulation pipeline that we use to generate Monte Carlo realizations of the data, which are needed for covariance matrix estimation and tests of the likelihood.  Section~\ref{sec:interpretation} describes the likelihood function that we use to interpret the data and presents constraints on cosmological parameters using our tSZ PDF measurements.  We discuss the results and conclude in Section~\ref{sec:outlook}.

We assume a flat $\Lambda$CDM cosmology throughout.  Unless stated otherwise, parameters are set to the WMAP9+eCMB+BAO+$H_0$ maximum-likelihood values~\cite{Hinshawetal2013}.  For brevity, we refer to this as the WMAP9 cosmology.  In particular, $\sigma_8 = 0.817$ is our fiducial value for the rms amplitude of linear density fluctuations on $8$ Mpc$/h$ scales at $z=0$.  This value is also consistent at the $1\sigma$ level with the initial cosmological results from \emph{Planck}~\cite{Planck2013params}.  All masses are quoted in units of $M_{\odot}/h$, where $h \equiv H_0/(100 \, \mathrm{km} \, \mathrm{s}^{-1} \, \mathrm{Mpc}^{-1})$ and $H_0$ is the Hubble parameter today.  All distances and wavenumbers are in comoving units of $\mathrm{Mpc}/h$.

\section{Data}
\label{sec:data}

The Atacama Cosmology Telescope~\cite{Fow07,Swe11,Dunneretal2013} is a six-meter telescope located on Cerro Toco in the Atacama Desert of northern Chile.  
In addition to the primordial CMB, ACT data~\cite{Dunneretal2013,Dasetal2014} contain the imprint of massive galaxy clusters seen through the tSZ effect.  A sample of high-significance clusters discovered in the ACT data are reported in~\cite{Marriageetal2011,Hasselfieldetal2013}, with additional optical and X-ray characterization described in~\cite{Menanteauetal2010,Menanteauetal2013}. These clusters are used to constrain cosmological parameters in~\cite{Hasselfieldetal2013}, partially based on dynamical masses measured in~\cite{Sifonetal2013}.  Further dynamical mass measurements are reported in~\cite{Kirketal2014}, while~\cite{Hiltonetal2013} investigates the clusters' stellar mass content and~\cite{Sehgaletal2013} investigates the tSZ -- optical richness scaling relation.

The analysis presented in this paper is a direct extension of the results presented in~W12, in which the tSZ signal in the ACT Equatorial map (from all groups and clusters in the field) is measured through its contribution to the skewness of the temperature distribution in the map.  
We work with a co-added map made from two seasons of observation at 148 GHz in the equatorial field in 2009 and 2010.  The map is comprised of six $3\degree \times 18\degree$ patches of sky.  The map noise level in CMB temperature units is $\approx 18\, \mu \mathrm{K}$ arcmin.  In contrast to~W12, we do not use any 218 GHz data in the primary analysis in this work.  
By restricting our analysis to negative temperature values in the 148 GHz histogram, we avoid essentially any contaminating CIB emission (this contamination motivated the use of the 218 GHz data in~W12).  
The maps used in this analysis are calibrated to WMAP using the methods described in~\cite{HajCal,Dasetal2014}.  To prevent ringing effects arising when we Fourier transform the maps, we apodize them with a mask that smoothly increases from zero to unity over $0.1\degree$ from the edge of the maps.

The processing of the 148 GHz map in this analysis is almost identical to that in~W12, with a few minor exceptions.  First, in order to upweight tSZ signal in the data over the noise from the atmosphere and CMB on large scales and the instrumental noise on small scales, we filter the maps in Fourier space using the Wiener filter shown in Fig.~1 of~W12, which was originally constructed by dividing the best-fit tSZ power spectrum from~\cite{Dunkleyetal2011} by the total average power spectrum measured in the data.  We consistently apply the same Fourier-space filter throughout this work to data, simulations, and theoretical calculations, unless explicitly stated.  In contrast to~W12, we do not remove signal in the $-100< \ell_{\mathrm{dec}}<100$ stripe along the Fourier axis corresponding to declination, a processing step previously required to avoid contamination by scan noise, which is no longer a significant issue in the final maps used here.  The beam uncertainties over the multipole range preserved by the filter are $<0.5$\%~\cite{Hasselfieldetal2013beam}.  After filtering, we mask pixels within $5.25$ arcmin of the edges of the maps in order to prevent any edge effects that could occur as a result of Fourier transforming, despite the apodization described above.  The edge mask removes $13.3$\% of the original map.

Second, as in~W12, we use template subtraction to remove SNR~$>5$ point sources from the map (primarily radio sources)~\cite{Mar11}.  In addition, we mask a circular region of radius 5 arcmin around the subtracted sources after applying the Fourier-space filter described in the previous paragraph, in order to reduce any ringing effects that might arise.  The point source mask removes $1.7$\% of the original map, corresponding to $\approx 260$ sources.  We verify that no non-Gaussian structure is produced in the histogram of simulated maps after point source subtraction, filtering, and masking.  
In total, $15$\% of the 148 GHz map is masked in the processing procedure, but the masked pixel locations are nearly independent of the tSZ field and should therefore not substantially alter the signal we are concerned with here.  Very massive clusters are not expected to lie under the SNR~$>5$ point sources in our maps, as their large tSZ decrements would prevent source detection.  Although some less massive clusters may be associated with the detected sources, these clusters are numerous and hence the small number of masked pixels is unlikely to affect our results.

\begin{figure}
\centering
\includegraphics[totalheight=0.5\textheight]{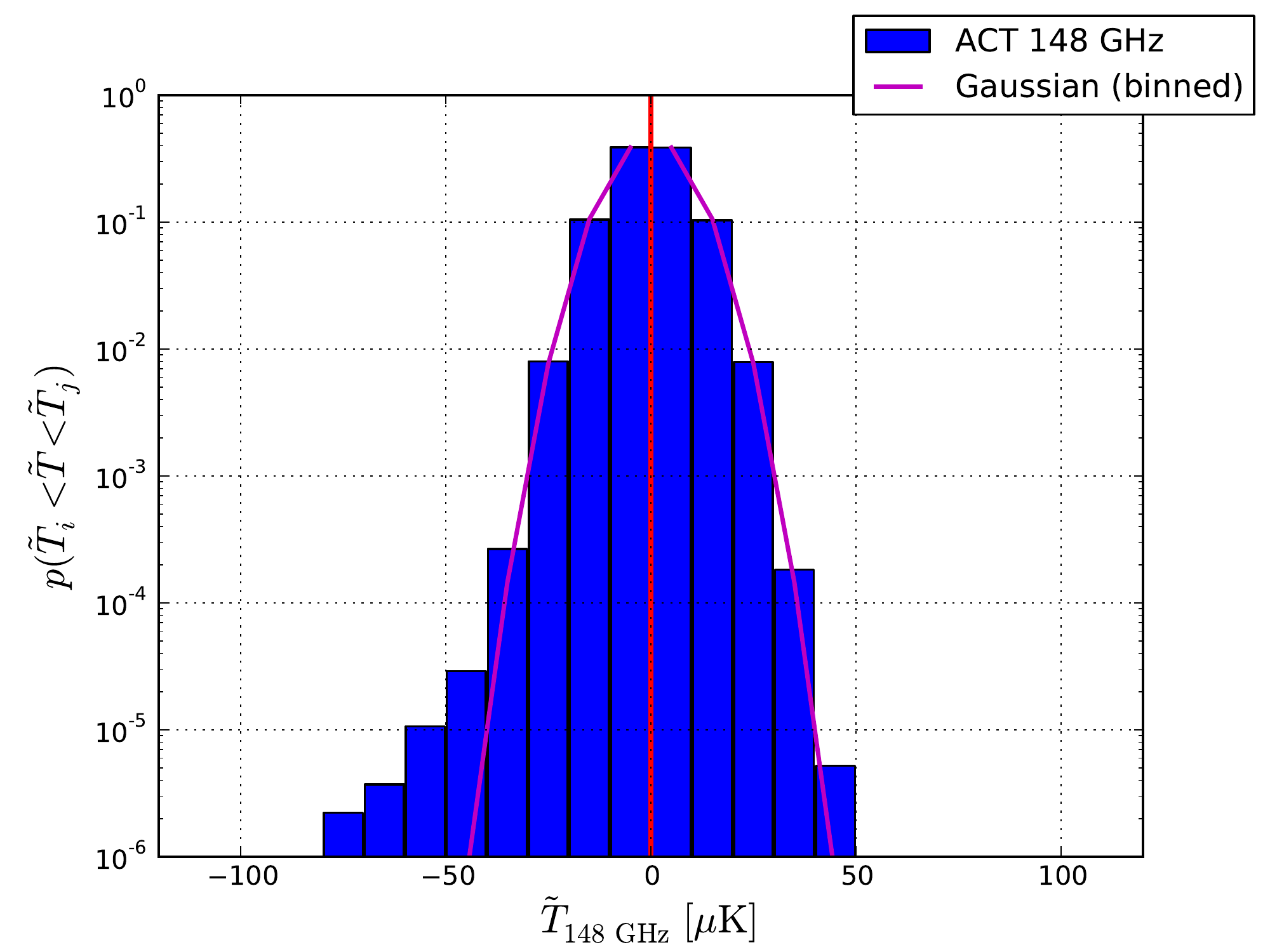}
\caption{Histogram of pixel temperature values in the ACT 148 GHz Equatorial CMB map after filtering and point source masking.  The solid red vertical line denotes $\tilde{T} = 0$.  The solid magenta curve is a Gaussian with variance given by the rms of the filtered, processed map; it is not a fit and is shown here only for visual reference.  The tSZ effect is responsible for the significant non-Gaussian tail on the negative side of the PDF ($\tilde{T} < 0$).  
\label{fig.ACTPDF}}
\end{figure}

Fig.~\ref{fig.ACTPDF} shows the binned PDF (i.e., histogram) of the pixel temperature values in the filtered, processed ACT Equatorial 148 GHz temperature map.  We use bins of width $10 \, \mu$K, extending to $\pm 120 \, \mu$K, well beyond where the negative tail of the distribution cuts off.  The number of bins is primarily motivated by the difficulty in computing the covariance matrix of the PDF for a large number of bins (see Sections~\ref{sec:covmat} and~\ref{sec:sims}).  Fig.~\ref{fig.ACTPDF} is equivalent to Fig.~2 in~W12, but with a coarser binning and slightly more pixels in the $\tilde{T} > 0$ tail of the distribution, since we have not applied the additional 218 GHz IR mask from~W12.  The mask is unnecessary because here we do not consider the positive side of the PDF ($\tilde{T} > 0$).  
For reference, we also show a Gaussian curve in Fig.~\ref{fig.ACTPDF}, with variance computed from the rms of the filtered, processed 148 GHz map.  A clear non-Gaussian excess is visible on the negative side of the histogram.

\begin{figure}
\centering
\includegraphics[totalheight=0.5\textheight]{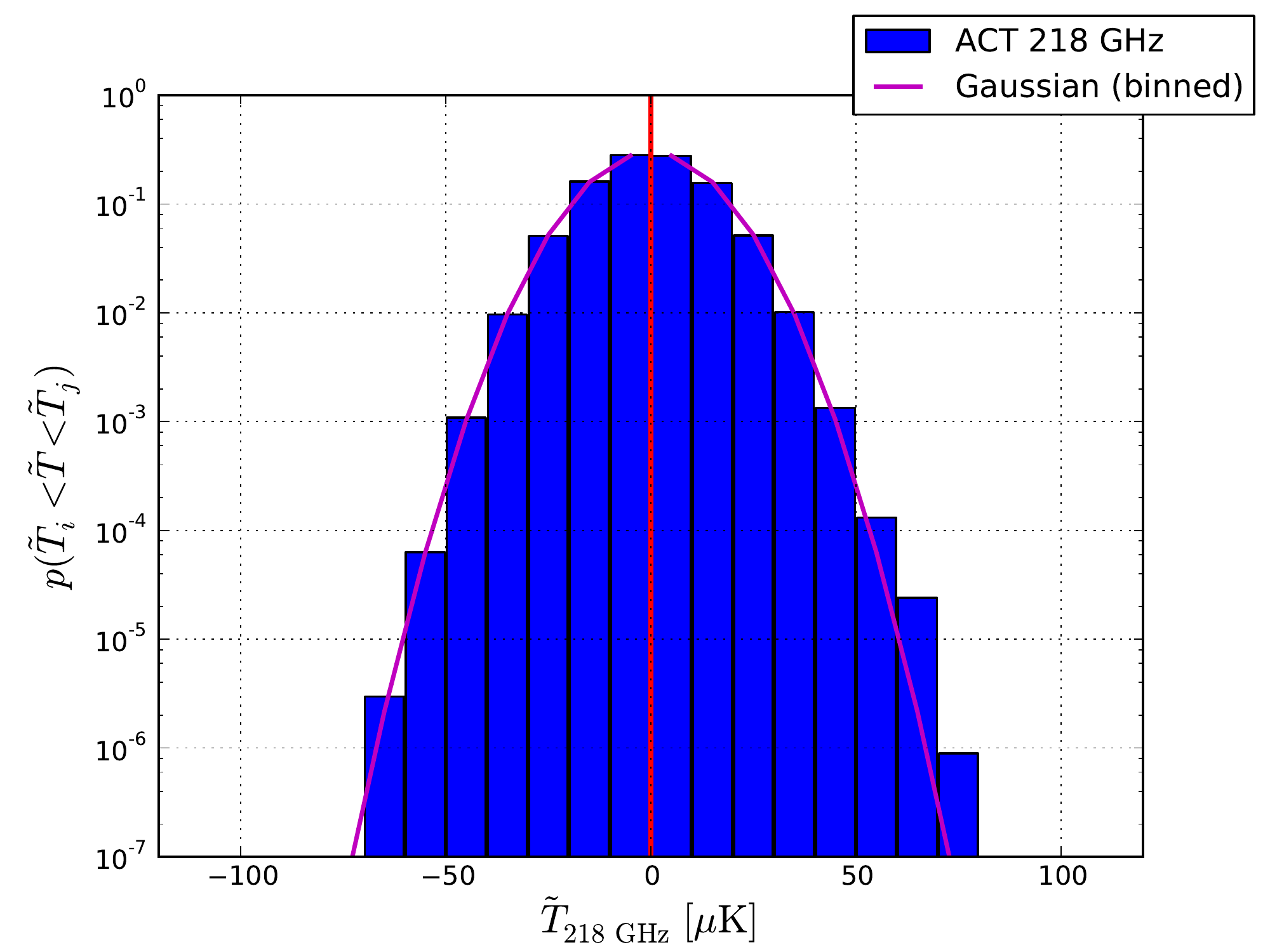}
\caption{Histogram of pixel temperature values in the ACT 218 GHz Equatorial CMB map after filtering and masking of point sources and Galactic dust.  The solid red vertical line denotes $\tilde{T} = 0$.  The solid magenta curve is a Gaussian with variance given by the rms of the filtered, processed map; it is not a fit.  In contrast to Fig.~\ref{fig.ACTPDF}, the negative side of the histogram has no non-Gaussian tail, which provides evidence that the feature in the 148 GHz histogram is due to the tSZ effect.  The excess on the positive side of the 218 GHz histogram is likely due to IR emission.
\label{fig.ACTPDF218}}
\end{figure}

To provide evidence that the negative non-Gaussian tail in the 148 GHz histogram arises from the tSZ effect, we process the ACT 218 GHz Equatorial map through the same pipeline after appropriate modifications to treat the different beams at the two frequencies.  We then compute the 218 GHz histogram.  The tSZ spectral function vanishes at 218 GHz, so no negative non-Gaussian tail should be seen at this frequency.  Due to the increased brightness of Galactic dust at 218 GHz compared to 148 GHz, we apply an additional dust mask based on Model 8 of~\cite{Finkbeineretal1999} (FDS) in the 218 GHz analysis.  We extrapolate the FDS map to 218 GHz and construct a mask that removes all pixels above a flux cut of $5.44$ MJy/sr, following~\cite{Dasetal2014}.  We confirm in the cosmological analysis of Section~\ref{sec:interpretation} that applying this Galactic dust mask to the 148 GHz maps does not change any of our results, with the best-fit $\sigma_8$ changing by less than one-third of a standard deviation.  
The histogram of 218 GHz temperature values resulting from this analysis is shown in Fig.~\ref{fig.ACTPDF218}.  The negative side of the histogram clearly has no significant non-Gaussian excess.

\section{Theory}
\label{sec:theory}
\subsection{Thermal SZ Effect}
\label{sec:theorymodel}
In the absence of relativistic corrections, the tSZ-induced change in the observed CMB temperature at frequency $\nu$ at angular position $\vec{\theta}$ on the sky with respect to the center of a cluster of mass $M$ at redshift $z$ is given by~\cite{Sunyaev-Zeldovich1970}:
\ba
\label{eq.tSZdef}
\frac{T_{\nu}(\vec{\theta}, M, z)}{\mathcal{T}_{\mathrm{CMB}} } & = & g_{\nu} y(\vec{\theta}, M, z) \nonumber\\
 & = & g_{\nu} \frac{\sigma_T}{m_e c^2} \int_{\mathrm{LOS}} P_e \left( \sqrt{l^2 + d_A^2 |\vec{\theta}|^2}, M, z \right) dl \,,
\ea
where $g_{\nu} = x\,\mathrm{coth}(x/2)-4$ is the tSZ spectral function with $x \equiv h\nu/(k_B \mathcal{T}_{\mathrm{CMB}})$, $y(\vec{\theta}, M, z)$ is the Compton-$y$ parameter, $\sigma_T$ is the Thomson scattering cross-section, $m_e$ is the electron mass, and $P_e(\vec{r},M,z)$ is the ICM electron pressure at (three-dimensional) separation $\vec{r}$ from the cluster center.  Our theoretical calculations assume that the pressure profile is spherically symmetric, i.e., $P_e(\vec{r},M,z) = P_e(r,M,z)$ where $r = |\vec{r}|$.  The integral in Eq.~(\ref{eq.tSZdef}) is computed along the line-of-sight such that $r^2 = l^2 + d_A^2(z) \theta^2$, where $l$ is the line-of-sight distance, $d_A(z)$ is the angular diameter distance to redshift $z$, and $\theta \equiv |\vec{\theta}|$.  We assume that the ICM consists of a fully ionized ideal gas of hydrogen and helium with primordial abundances.  Ion--electron equilibration implies that the electron pressure $P_e(\vec{r},M,z)$ is related to the thermal gas pressure via $P_{th} = P_e (5 X_H+3)/(2(X_H+1)) = 1.932 P_e$, where $X_H=0.76$ is the primordial hydrogen mass fraction.

Eq.~(\ref{eq.tSZdef}) is only valid in the non-relativistic limit, i.e., when the electron temperature $\mathcal{T}_e$ satisfies~$k_B \mathcal{T}_e \ll m_e c^2$.  In this case, $g_{\nu}$ is independent of $\mathcal{T}_e$ as written above.  Integrating the non-relativistic $g_{\nu}$ over the ACT $148$ GHz bandpass yields an effective frequency of $146.9$ GHz~\cite{Swe11}.  Unless otherwise specified, we compute all tSZ quantities at this frequency throughout the rest of the paper, although quantities may be labeled with ``$148$ GHz.''  We will also often abbreviate $T \equiv T_{148\, {\rm GHz}}$.

For our analysis, relativistic corrections are small but non-negligible.  To calculate these corrections, we work to second order in $k_B \mathcal{T}_e/(m_e c^2)$ using the results of~\cite{Nozawaetal2006}.  In this case, the spectral function depends on both frequency and electron temperature, i.e., $g_{\nu}$ in Eq.~(\ref{eq.tSZdef}) becomes $g_{\nu}(\mathcal{T}_e)$.  We compute $\mathcal{T}_e$ as a function of mass and redshift using the $\Delta = 200$ scaling relation of~\cite{Arnaudetal2005}.  Although this relation comes with some systematic uncertainty, it yields a subdominant correction to the overall relativistic correction to the tSZ signal, which is $\approx 2-12$\% for the clusters in our calculations (see below for details on the mass and redshift limits).  The sign of the relativistic correction is such that a decrement at $148$ GHz becomes less negative (i.e., it is ``filled in'').  Because we include relativistic corrections, the tSZ results cannot be phrased simply in terms of Compton-$y$ and are specific to our effective frequency of $146.9$ GHz.

Our definition of the virial mass $M$~\cite{Bryan-Norman1998}, convention for the concentration-mass relation~\cite{Duffyetal2008}, and fiducial ICM pressure profile model~\cite{Battagliaetal2012} are the same as in~\cite{Hill-Pajer2013,Hill-Spergel2014}.  The pressure profile model is a parametrized fit to stacked profiles extracted from the ``AGN feedback'' simulations described in~\cite{Battagliaetal2010}.  The model fully describes the pressure profile of the ICM gas as a function of $M_{200}$ and redshift, where $M_{\Delta}$ is the mass enclosed within a radius $r_{\Delta}$ such that the mean enclosed density is $\Delta$ times the critical density at the cluster redshift.  We convert between mass definitions when needed using a Navarro-Frenk-White density profile~\cite{NFW1997} and the concentration-mass relation~\cite{Duffyetal2008}.  Unless otherwise specified (e.g., as $M_{200}$ or $M_{500}$), all masses quoted in the paper refer to the virial mass as defined in~\cite{Bryan-Norman1998}.  The pressure profile model and the halo mass function are the only necessary inputs for our analytic calculations of the tSZ PDF (see Section~\ref{sec:tSZonlyPDF}).

The ``AGN feedback'' simulations include virial shock heating, radiative cooling, and sub-grid prescriptions for star formation and feedback from supernovae and active galactic nuclei (AGN)~\cite{Battagliaetal2010}.  Non-thermal pressure support due to bulk motion and turbulence in the ICM outskirts are captured by the smoothed particle hydrodynamics (GADGET-2) method used for the simulations.  The pressure profile extracted from these simulations has been found to agree with many recent X-ray and tSZ measurements~\cite{Arnaudetal2010,Sunetal2011,Planck2013stack,Planck2012Coma,Planck2013ymap,Hajianetal2013,Hill-Spergel2014}.  In this model, the ICM thermal pressure profile is parametrized by a dimensionless ``generalized Navarro-Frenk-White'' form, which has been used in many observational and numerical studies~(e.g.,~\cite{Nagaietal2007,Arnaudetal2010,Plaggeetal2010,Planck2013stack}):
\be
\label{eq.GNFW}
\frac{P_{th}(x)}{P_{200}} = \frac{\Pi_0 \left( x/x_c\right)^{\gamma}}{\left(1+\left( x/x_c \right)^{\alpha} \right)^{\beta}} \,\,\,\,\,, x \equiv r/r_{200} \,,
\ee
where $x$ is the dimensionless distance from the cluster center, $x_c$ is a core scale length, $\Pi_0$ is a dimensionless amplitude, $\alpha$, $\beta$, and $\gamma$ describe the logarithmic slope of the profile at intermediate ($x \approx x_c$), large ($x \gg x_c$), and small ($x \ll x_c$) radii, respectively, and $P_{200}$ is the self-similar amplitude for thermal gas pressure at $r_{200}$~\cite{Kaiser1986,Voit2005}:
\be
\label{eq.P200c}
P_{200} = \frac{200 \, G M_{200} \rho_{c}(z) \Omega_b}{2\, \Omega_m r_{200}} \,.
\ee
This functional form is fit to the stacked pressure profiles of clusters extracted from the simulations described above as a function of mass and redshift.  The mass and redshift dependences of these parameters capture departures from self-similarity.  The results are as follows~\cite{Battagliaetal2012} (note that $\alpha=1.0$ and $\gamma=-0.3$ are held fixed due to parameter degeneracies)\footnote{N.B. The denominator of the mass-dependent factor in these expressions has units of $M_{\odot}$ rather than $M_{\odot}/h$ as used elsewhere in this paper.}:
\beqn
\Pi_0(M_{200},z) & = & 18.1 \left( \frac{M_{200}}{10^{14} \,\, M_{\odot}} \right)^{0.154} \left( 1+z \right)^{-0.758} \label{eq.Pi0batt} \\
x_c(M_{200},z) & = & 0.497 \left( \frac{M_{200}}{10^{14} \,\, M_{\odot}} \right)^{-0.00865} \left( 1+z \right)^{0.731} \label{eq.xcbatt} \\
\beta(M_{200},z) & = & 4.35 \left( \frac{M_{200}}{10^{14} \,\, M_{\odot}} \right)^{0.0393} \left( 1+z \right)^{0.415} \label{eq.betabatt} \,.
\eeqn
Eqs.~(\ref{eq.GNFW})--(\ref{eq.betabatt}) fully describe the ICM thermal pressure profile as a function of mass, redshift, and cluster-centric radius.

Although this model agrees with a number of X-ray and tSZ measurements, the pressure--mass relation remains a significant source of uncertainty in current cosmological constraints from tSZ measurements (e.g.,~\cite{Hasselfieldetal2013,Reichardtetal2013,Planck2013counts,Wilsonetal2012,Crawfordetal2014,Hill-Spergel2014}).  In this work, we parametrize the uncertainty in terms of $P_0$, the overall normalization of the pressure--mass relation.  Allowing $P_0$ to vary, we thus slightly modify Eq.~(\ref{eq.Pi0batt}):
\be
\label{eq.P0def}
\Pi_0(M_{200},z) = P_0 \left( 18.1 \left( \frac{M_{200}}{10^{14} \,\, M_{\odot}} \right)^{0.154} \left( 1+z \right)^{-0.758} \right) \,.
\ee
The fiducial value $P_0=1$ therefore corresponds to the unmodified pressure profile from~\cite{Battagliaetal2012}.  This approach is equivalent to parameterizing the ICM uncertainty via $(1-b)$, the ``hydrostatic mass bias,'' a free parameter that sets the overall normalization of the ``universal pressure profile'' from~\cite{Arnaudetal2010}, which is derived using hydrostatic X-ray masses.  Our fiducial ``AGN feedback'' model ($P_0 = 1$) corresponds to a hydrostatic mass bias $\approx 5$--$15$\% (i.e., $(1-b) \approx 0.85$--$0.95$) for the massive, low-redshift population of clusters studied in~\cite{Arnaudetal2010}, though this value varies with cluster-centric radius (see Fig.~2 of~\cite{Battagliaetal2010}).  The pressure--mass normalization (or hydrostatic mass bias) is a quantity averaged over all clusters in the population under study; it is expected to be a function of cluster mass and redshift, and likely to possess scatter about its mean, but to lowest order we aim to constrain its mean value over the cluster population.  Thus, $P_0 = 1$ corresponds to the fiducial ICM pressure profile model, whereas $P_0 < 1$ ($> 1$) implies a typical thermal pressure less (greater) than that expected in the fiducial model, corresponding to a larger (smaller) value of the hydrostatic mass bias than that already present in the fiducial model.  The final result of Eqs.~(\ref{eq.tSZdef})--(\ref{eq.P0def}) is a prescription for the tSZ temperature fluctuation, $T(\theta, M, z)$.  

In all of the relevant calculations for the ACT analysis, an $\ell$-space Wiener filter is used to increase the tSZ signal-to-noise in the data.  The filter is identical to that used in~W12.  In addition, we use temperature maps in which the beam has not been deconvolved in order to prevent pixel-to-pixel noise correlations in the final maps as much as possible.  We include these effects in $\ell$-space in the analytic calculations by Fourier transforming $T(\theta, M, z)$ for each cluster, multiplying by the filter and beam functions, and then Fourier transforming back to real space.  The pixel window function is so close to unity over the relevant multipole range that we need not consider it.  
From this point onward, $\tilde{T}(\theta, M, z)$ refers to the filter- and beam-convolved $T$-profile for each cluster of mass $M$ and redshift $z$.

\subsection{Noiseless tSZ-Only PDF}
\label{sec:tSZonlyPDF}

The most straightforward and accurate way to compute the tSZ PDF is using cosmological hydrodynamics simulations, an approach which was pursued by several groups roughly a decade ago~\cite{daSilvaetal2000,Seljaketal2001,Springeletal2001,Springeletal2001ERR,Refregier-Teyssier2002,Zhangetal2002}.  However, it is computationally intractable to run such simulations for a variety of cosmological parameter values.  Thus, in order to constrain cosmology using the tSZ PDF, a less computationally taxing, analytic method is required.  
For this purpose, we develop a halo model-based approach to calculate the tSZ PDF.  This method requires some assumptions, as we describe below, but it does not assume that the non-Gaussianity in the tSZ field is weak, as has been assumed in earlier approaches~(e.g.,~\cite{Cooray2000}).  We validate the accuracy of the model through a comparison to hydrodynamical simulations.

The value of the tSZ PDF integrated over a bin $b_i \equiv (\tilde{T}_i, \tilde{T}_{i+1})$, which we denote by $p(\tilde{T}_{i} < \tilde{T} < \tilde{T}_{i+1})$ (where $\tilde{T}$ is the filtered $146.9$ GHz tSZ temperature decrement) or simply $\langle p_i \rangle$, corresponds to the expected fraction of the sky subtended by $\tilde{T}$ values within that bin.  Thus, a straightforward way to compute the signal is to calculate the relevant fraction of sky contributed by each cluster of mass $M$ and redshift $z$, and then sum over all the clusters in the universe.  Let $f(\tilde{T},M,z) d\tilde{T}$ be the sky area subtended by the $\tilde{T}$-profile of a cluster of mass $M$ and redshift $z$ in the range $(\tilde{T},\tilde{T}+d\tilde{T})$.  Define $\theta(\tilde{T},M,z)$ to be the inverse function of the filtered temperature fluctuation profile $\tilde{T}(\theta,M,z)$ in Eq.~(\ref{eq.tSZdef}).  Then $f(\tilde{T},M,z) d\tilde{T}$ is given by
\beq
f(\tilde{T},M,z) d\tilde{T} = 2 \pi \theta(\tilde{T},M,z) \left. \frac{d\theta}{d\tilde{T}} \right|_{\tilde{T},M,z} d\tilde{T} \,.
\label{eq.fofy}
\eeq
The noiseless tSZ PDF $\langle p_i \rangle_{\rm noiseless}$ in bin $b_i$ is then given by summing up the contributions to this bin from all clusters in the universe:
\beqn
\langle p_i \rangle_{\rm noiseless} & = & \int dz \frac{d^2 V}{dz d\Omega} \int dM \frac{dn}{dM} \int_{\tilde{T}_i}^{\tilde{T}_{i+1}} d\tilde{T} f(\tilde{T},M,z) \nonumber \\
    & = & \int dz \frac{d^2 V}{dz d\Omega} \int dM \frac{dn}{dM} \pi \left( \theta^2(\tilde{T}_{i+1},M,z) - \theta^2(\tilde{T}_{i},M,z) \right) \,,
\label{eq.yPDFnonoise}
\eeqn
where $d^2 V/dz d\Omega = c \chi^2(z)/H(z)$ is the comoving volume per steradian per unit redshift in the assumed flat cosmology, $\chi(z)$ is the comoving distance to redshift $z$, $H(z)$ is the Hubble parameter at redshift $z$, and $dn(M,z)/dM$ is the halo mass function (the number of halos at redshift $z$ per unit comoving volume and per unit mass), for which we use the prescription of~\cite{Tinkeretal2008}.  The PDF is normalized such that $\Sigma_i \langle p_i \rangle = 1$, which requires that a contribution $(1-F_{\rm clust})$ should be added to Eq.~(\ref{eq.yPDFnonoise}) if bin $b_i$ includes $\tilde{T} = 0$, where $F_{\rm clust}$ is the total sky fraction subtended by clusters (see below) --- this contribution accounts for the fraction of sky possessing no tSZ signal.  Since most of the sky is not subtended by galaxy clusters (i.e., $F_{\rm clust} < 1$), the tSZ PDF is sharply peaked at $\tilde{T} = 0$ in the noiseless, tSZ-only case; for the noise-convolved, observable case (see Section~\ref{sec:noisyPDF}), this peak is smoothed out by the presence of CMB fluctuations, instrumental noise, and other contributions to the microwave sky.

The primary assumption in our halo model-based approach is the neglect of any effects due to overlapping clusters along the line-of-sight.  This approximation breaks down if one na\"{i}vely extends the computation down to arbitrarily low masses or assigns an arbitrarily large outer boundary for each cluster.  The validity of the approximation is encoded in the na\"{i}ve sky fraction subtended by all clusters in the calculation, assuming no overlaps:
\beq
F_{\rm clust} = \int dz \frac{d^2V}{dzd\Omega} \int dM \frac{dn}{dM} \pi \theta_{\rm out}^2(M,z) \,,
\label{eq.Fclust}
\eeq
where $\theta_{\rm out}(M,z) \equiv r_{\rm out}(M,z)/d_A(z)$ is the outer angular boundary of a cluster of mass $M$ at redshift $z$.  We take the outer boundary to be twice the virial radius (as defined in~\cite{Hill-Pajer2013,Hill-Spergel2014}), which is roughly the location of the accretion shock seen in hydrodynamical simulations of cluster formation~(e.g.,~\cite{Molnaretal2009}).  The results of our analysis of the ACT tSZ PDF are not sensitive to this choice because the $\tilde{T}$-values in the outer regions of clusters are so close to zero that they contribute only to the noise-dominated regime of the measured PDF (see Section~\ref{sec:noisyPDF}).  The approximation of neglecting line-of-sight overlaps is valid as long as $F_{\rm clust} \ll 1$.  Numerical testing using the mass function of~\cite{Tinkeretal2008} indicates that this approximation breaks down when integrating the mass function down to a mass scale $\approx 10^{14} M_{\odot}/h$.  
Thus, we take the mass limits of the integrals in all expressions in this work to be $2 \times 10^{14} M_{\odot}/h < M < 5 \times 10^{15} M_{\odot}/h$, which give $F_{\rm clust} \approx 0.3$ (we verify with simulations in Section~\ref{sec:sims} that overlap effects remain negligible for these mass limits).  The redshift integration limits are $0.005 < z < 3$, where the lower limit is chosen to prevent an apparent divergence at $z=0$, which is an artifact of the flat-sky approximation in Eq.~(\ref{eq.yPDFnonoise}) (the flat-sky approximation is only taken over the angular size of the cluster).  All computations are converged using these limits. With the noise and beams appropriate for ACT, clusters at or below the lower mass limit are simply absorbed in the noise-dominated region of the PDF (see Section~\ref{sec:noisyPDF}).  However, due to this approximation, a direct comparison of our analytic results to a zero-noise, tSZ-only map extracted from a hydrodynamical simulation will be expected to disagree somewhat in the low-$|\tilde{T}|$ regime, since the analytic calculation does not include all of the low-$|\tilde{T}|$ signal by construction.  For the physically observable PDF, the ACT noise and beams are large enough that the data are not sensitive to $\tilde{T}$ values sourced by clusters below our lower mass cutoff.

\begin{figure}
\centering
\includegraphics[totalheight=0.5\textheight]{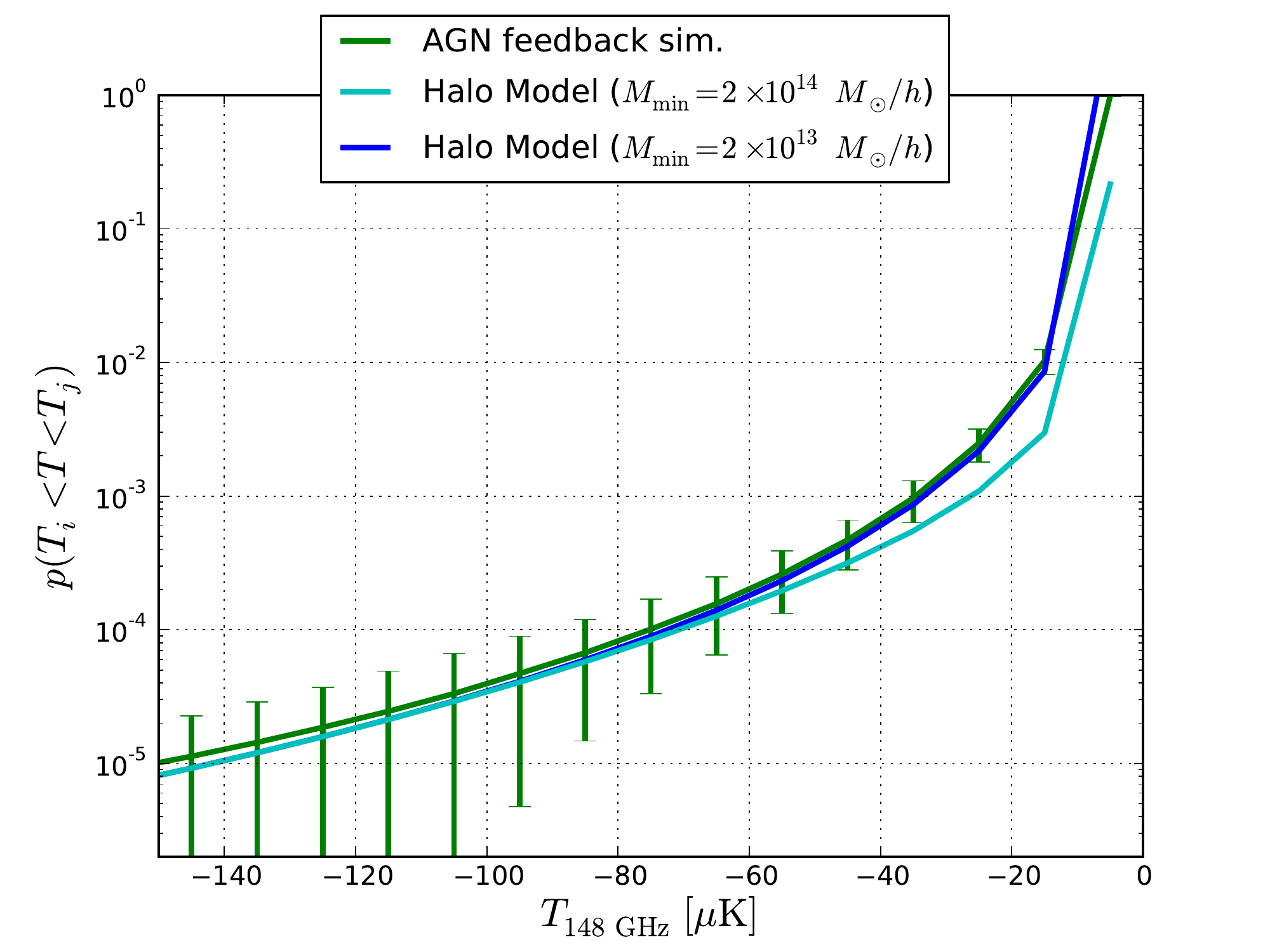}
\caption{Comparison of the noiseless halo model computation of the tSZ PDF using Eq.~(\ref{eq.yPDFnonoise}) to the PDF directly measured from the cosmological hydrodynamics simulations of~\cite{Battagliaetal2010}.  The error bars are computed from the scatter amongst the 390 maps extracted from the simulation.  No filtering or noise convolution has been performed for these calculations, and thus the $T$ values cannot be directly compared to those shown in the other plots in the paper.  The comparison demonstrates that the halo model approach works well, except for discrepancies at small-$|T|$ values arising from the breakdown of the halo model assumptions, which do not affect the fully noise-convolved ACT PDF calculations (see Section~\ref{sec:noisyPDF}).
\label{fig.hydrosimcomp}}
\end{figure}

We validate the halo model-based theory described above by comparing to the tSZ PDF directly extracted from cosmological hydrodynamics simulations.  We use the simulations from which our fiducial pressure profile model was extracted~\cite{Battagliaetal2010} in order to facilitate a like-for-like comparison, extracting 390 Compton-$y$ maps of area $(4.09\degree)^2$ each.  Relativistic corrections are not included in the construction of the maps (they are pure Compton-$y$~maps), and thus for this exercise alone we neglect relativistic corrections in our analytic calculations.  The maps are direct line-of-sight projections of randomly rotated and translated simulation volumes, with an upper redshift cut at $z=1$ in order to decrease correlations due to common high-redshift objects in the extracted maps.  The same redshift cut is applied to the theoretical calculations for consistency (only for this particular exercise).  We consider two values for the lower mass cutoff in Eq.~(\ref{eq.yPDFnonoise}): the fiducial $2 \times 10^{14} M_{\odot}/h$ and a lower value of $2 \times 10^{13} M_{\odot}/h$.  This particular calculation uses the same values for the cosmological parameters as those used in the simulations (see~\cite{Battagliaetal2010}), which differ from our fiducial WMAP9 parameters.  
Also, we do not include the Wiener filter or noise in this calculation, and thus the resulting $T$ values are not directly comparable to those found elsewhere in the paper (which are labeled $\tilde{T}$ for clarity).

The results of the analysis are shown in Fig.~\ref{fig.hydrosimcomp}.  Overall, the agreement between the halo model and the simulations is excellent.  At the largest $|T|$-values, the slight discrepancy is due to an excess of massive clusters in the simulations of~\cite{Battagliaetal2010} compared to the Tinker mass function (see the appendix of~\cite{Battagliaetal2012}).  At the smallest $|T|$-values, discrepancies emerge due to the breakdown of our assumption that clusters do not overlap along the line-of-sight --- this is responsible for the unphysical value of $p>1$ in the lowest bin for the $M_{\rm min} = 2 \times 10^{13} M_{\odot}/h$ calculation.  One can also see that the lowest bins receive some contributions from halos with masses between $2 \times 10^{13} M_{\odot}/h$ and $2 \times 10^{14} M_{\odot}/h$; however, we will demonstrate in the following section that the tSZ signal from all such objects is subsumed into the noise in the ACT data.  This noiseless comparison is simply a test of the halo model framework.

The general agreement seen in Fig.~\ref{fig.hydrosimcomp} supports three assumptions made in the halo model approach.  First, the comparison verifies that neglecting scatter in the $P_e(r,M,z)$ relation does not invalidate our analytic theory; this scatter is obviously present in the hydrodynamical simulations, which agree well with our results computed using only the mean $P_e(r,M,z)$ relation from~\cite{Battagliaetal2012}.  Second, the comparison indicates that tSZ signal from non-virialized regions is negligible for our analysis --- this signal is present in the maps from the simulations (which are direct line-of-sight integrations of the simulation volume), but neglected in the halo model approach.  The halo model calculations, when extended down to low enough masses, appear to account for essentially all of the signal in the simulations.  Finally, the comparison verifies that the halo model's neglect of effects due to overlapping clusters along the line-of-sight is safe for a minimum mass cutoff satisfying the $F_{\rm clust} \ll 1$ criterion described above (we test this assumption further in Section~\ref{sec:sims}).  

\subsection{Observable tSZ PDF}
\label{sec:noisyPDF}

We now describe the general case, including non-tSZ contributions to the microwave sky, followed by the details of each non-tSZ component's contribution to the noise in the observed ACT PDF.  Let $\rho_i(\tilde{T})$ be the probability of observing a signal in bin $b_i \equiv (\tilde{T}_i, \tilde{T}_{i+1})$ given an input (physical) signal $\tilde{T}$:
\beq
\rho_i(\tilde{T}) = \int_{\tilde{T}_{i}}^{\tilde{T}_{i+1}} d\tilde{T}' N(\tilde{T}-\tilde{T}') \,,
\label{eq.noisekernel}
\eeq
where $N(\tilde{T}-\tilde{T}')$ is the noise PDF, which has units of inverse temperature.  The contribution of a single cluster to bin $b_i$ after including noise is then given by:
\beqn
g_i(M,z) & = & \int d\tilde{T} f(\tilde{T},M,z) \rho_i(\tilde{T}) \nonumber \\
                & = & \int d\theta \,\, 2 \pi \theta \, \, \rho_i \left( \tilde{T}(\theta,M,z) \right) \,,
\label{eq.gbdef}
\eeqn
where $\rho_i(\tilde{T})$ is defined by Eq.~(\ref{eq.noisekernel}), $f(\tilde{T},M,z)$ is defined by Eq.~(\ref{eq.fofy}), and the second line is a computationally simpler way of expressing the same quantity.  The results in the noiseless case discussed in Section~\ref{sec:tSZonlyPDF} can be recovered by taking $\rho_i(\tilde{T}) = \Theta(\tilde{T}-\tilde{T}_{i}) - \Theta(\tilde{T}-\tilde{T}_{i+1}) $, where $\Theta(x)$ is the Heaviside step function.  Adding the contribution from regions of the sky with zero intrinsic $\tilde{T}$-signal that fluctuate (due to noise) into bin $b_i$,  we obtain the final expression for the one-point PDF of the noise-convolved, observable $\tilde{T}$ field:
\beq
\langle p_i \rangle = (1-F_{\rm clust}) \rho_i(0) + \int dz \frac{d^2 V}{dz d\Omega} \int dM \frac{dn}{dM} \, g_i(M,z)\,,
\label{eq.yPDF}
\eeq
where the first term represents the contributions from noise fluctuations in intrinsically $\tilde{T} = 0$ pixels and the second term represents the physical tSZ contributions after noise convolution.  Eq.~(\ref{eq.yPDF}) has the convenient property that adding or subtracting low-mass clusters in the calculation of $F_{\rm clust}$ or the integral in the second term does not change the result, as this procedure simply shifts these contributions between either of the two terms.  Similarly, slight changes to the definition of the outer boundary of a cluster do not matter, as these regions are low-$|\tilde{T}|$ and thus subsumed into the noise-dominated region of the PDF (the same statement holds for changes in the upper redshift integration limit).  These statements are contingent upon the approximations described in Section~\ref{sec:tSZonlyPDF}, in particular that cluster overlaps along the line-of-sight can be neglected.  More precisely, we must be in the regime where $F_{\rm clust} \ll 1$.  As mentioned earlier, the approximations hold for our ACT analysis calculations with mass limits given by $\left\{2 \times 10^{14} M_{\odot}/h, 5 \times 10^{15} M_{\odot}/h \right\}$ and $r_{\rm out} = 2 r_{\rm vir}$, which give $F_{\rm clust} \approx 0.3$.  Although this value approaches the regime where the approximations break down, we verify directly using simulations in Section~\ref{sec:sims} that the analytic theory described here is valid for these mass and redshift limits.

We now detail the construction of the noise kernel $N(\tilde{T}-\tilde{T}')$.  In the case of homogeneous, uncorrelated, Gaussian noise, the noise kernel is
\beq
N_{G}(\tilde{T}-\tilde{T}') = \frac{1}{\sqrt{2\pi \sigma_N^2}} e^{-(\tilde{T}-\tilde{T}')^2/(2\sigma_N^2)} \,,
\label{eq.noisekernelgauss}
\eeq
where $\sigma_N$ is the single-pixel rms noise.  The homogeneous, Gaussian approximation is reasonable for computing forecasts, but in the ACT data analysis a more detailed treatment of the noise is required.  In particular, for ACT we must account for inhomogeneities in the instrumental and atmospheric noise contributions, which lead to a slightly non-Gaussian noise PDF.  In addition, we include a Gaussian noise contribution from the primordial CMB fluctuations (although this component is significantly reduced by the $\ell$-space filter used in the analysis).  Other noise contributions come from unresolved point sources and the CIB sourced by dusty star-forming galaxies --- however, we largely mitigate these sources of noise by considering only the $\tilde{T} < 0$ region of the ACT 148 GHz PDF, as described in Section~\ref{sec:data}.  Our procedure for constructing the noise kernel $N(\tilde{T}-\tilde{T}')$ appropriate for the ACT analysis is as follows:

\begin{itemize}

\item Instrumental and atmospheric noise: this component provides the largest contribution to the total noise in our filtered map.  Instead of assuming it to be Gaussian, we measure the instrumental and atmospheric noise directly using null maps constructed from the differences between single-season ``split'' maps.  The null maps are processed in exactly the same manner as the 148 GHz data map, but contain effectively no cosmological or astrophysical signal.  Thus they provide an accurate characterization of the instrumental and atmospheric noise.  We fit a cubic spline to this noise PDF, which we will refer to as $N_{\rm inst}(x)$.  Its contribution to the rms of the filtered map is $\sigma_{\rm inst} = 5.78 \, \mu{\rm K}$.  Although it is almost perfectly symmetric about $\tilde{T} = 0$ (and thus does not affect our earlier tSZ skewness results in~W12), the noise PDF has non-Gaussian tails that must be accounted for in this analysis.  These tails arise from atmospheric effects and from the slightly varying (Gaussian) noise level across the map.

\item CMB: the primary CMB anisotropies provide the second-largest contribution to the noise budget.  As discussed in Section~\ref{sec:interpretation}, their effects are marginalized over in the analysis, but to set the fiducial level we compute the CMB power spectrum $C_{\ell}$ for our fiducial WMAP9 cosmology using CAMB\footnote{{\tt http://camb.info}} and then obtain the corresponding variance via
\beq
\sigma_{\rm CMB}^2 = \sum_{\ell} \frac{2\ell+1}{4\pi} C_{\ell} F_{\ell}^2 b_{\ell}^2,
\label{eq.sumCell}
\eeq
where $F_{\ell}$ is the filter introduced in Section~\ref{sec:data} and $b_{\ell}$ is the ACT 148 GHz beam~\cite{Hasselfieldetal2013beam}.  This computation yields $\sigma_{\rm CMB} = 5.22 \, \mu{\rm K}$.  We treat the CMB PDF $N_{\rm CMB}(x)$ as a Gaussian with this variance.

\begin{figure}
\centering
\includegraphics[totalheight=0.5\textheight]{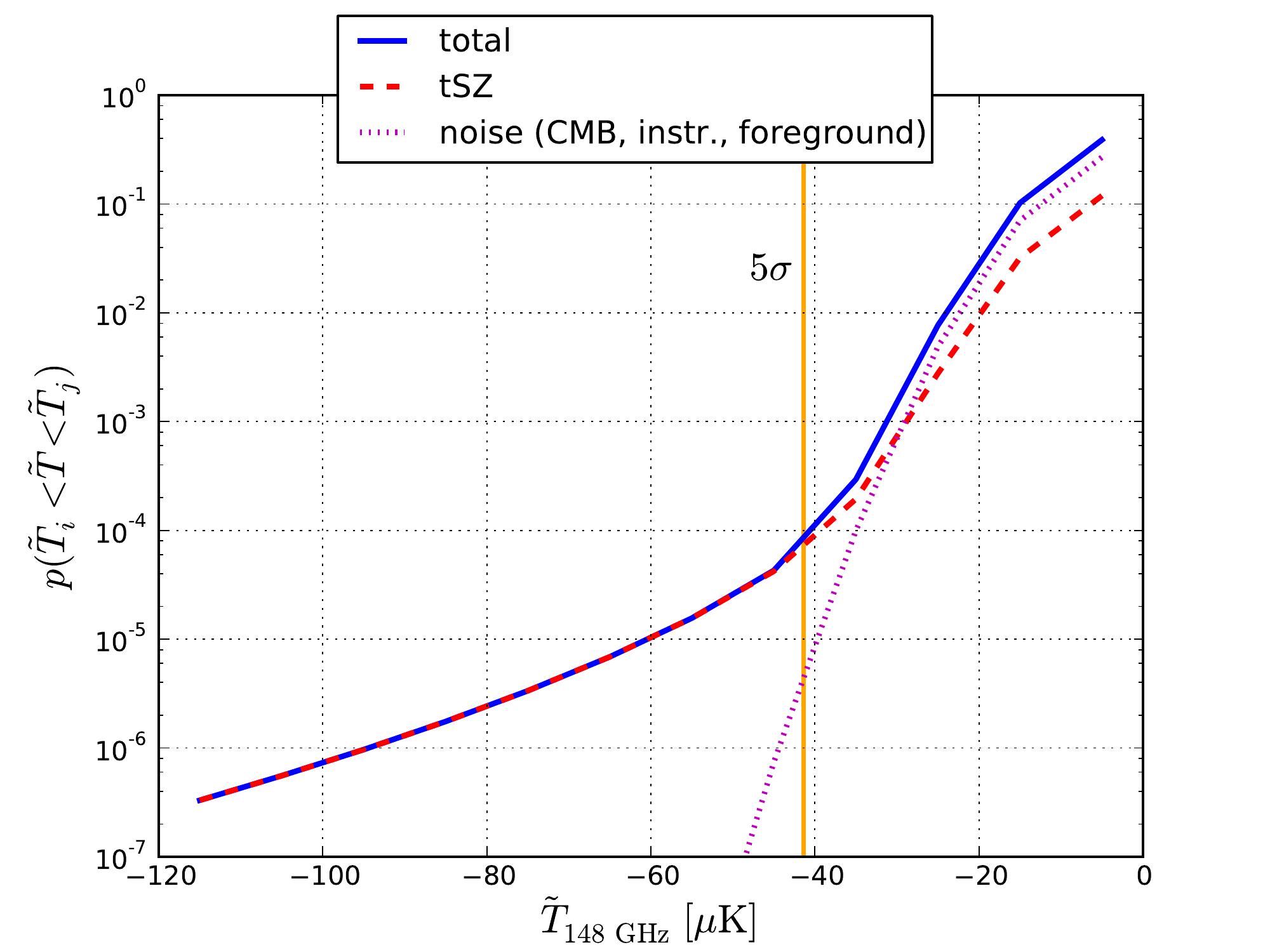}
\caption{The noise-convolved, observable tSZ PDF computed for our fiducial cosmology and pressure profile model using the analytic halo model framework described in Section~\ref{sec:noisyPDF}, with noise properties appropriate for the ACT Equatorial map.  The dashed red curve shows the tSZ contribution given by the second term in Eq.~(\ref{eq.yPDF}), the dotted magenta curve shows the noise contribution given by the first term in Eq.~(\ref{eq.yPDF}), and the solid blue line shows the total signal.  The vertical orange line corresponds to a $5\sigma$ temperature decrement in the ACT Equatorial map, where $\sigma = 8.28 \, \mu$K is the rms of the filtered map.  The PDF thus includes contributions from clusters well above the threshold for detection ($\gtrsim 5 \sigma$) as well as groups and clusters that comprise $\approx 3$--$4\sigma$ excursions in the map.  See Fig.~\ref{fig.masscontrib} for detailed characterization of the mass and redshift contributions to the signal, as well as Fig.~\ref{fig.maskHass} for contributions to the observed ACT PDF from individually detected clusters as a function of their SNR.
\label{fig.PDFfid}}
\end{figure}

\item Other contaminants: other sources of power in the microwave sky also contribute to the PDF in the ACT 148 GHz map.  The most important of these components are the CIB and emission from radio and infrared point sources.  Although our source masking procedure greatly reduces the IR and radio contributions to the PDF, an unresolved component (including Poisson and clustered terms) still remains in the map.  The CIB and source contributions are restricted to the $T > 0$ region of the PDF in the absence of $\ell$-space filtering, but leak slightly into the $\tilde{T} < 0$ region after we filter out the zero-mode of the map.  Excluding the $\tilde{T} > 0$ region removes nearly all of the CIB and point source contributions in our analysis.  We also discard the first bin in the $\tilde{T} < 0$ region to further ensure that no CIB or point source emission affects our results (see Section~\ref{sec:interpretation}).  This bin corresponds to $[-10\, \mu {\rm K}, 0\, \mu {\rm K}]$; we confirm using simulations~\cite{Sehgaletal2010} that all predicted CIB and point source leakage lies within this bin.  However, all non-tSZ sources still contribute to the noise with which the underlying tSZ PDF must be convolved; we model their contribution as a Gaussian with variance determined from the best-fit foreground parameters for the ACT Equatorial map in~\cite{Sieversetal2013}.  In particular, we sum the best-fit results from~\cite{Sieversetal2013} for all of the 148 GHz power spectra sourced by foregrounds except tSZ (which is not a foreground for our analysis).  Although some individual components are not tightly constrained (e.g., the kSZ power), the total non-tSZ power is well determined.  We then compute the resulting variance in our filtered map $\sigma_{\rm fg}^2$ using Eq.~(\ref{eq.sumCell}).  We find $\sigma_{\rm fg} = 2.33 \, \mu{\rm K}$.  We treat the foregrounds' PDF $N_{\rm fg}(x)$ as a Gaussian with this variance.  Since the CMB is also Gaussian, we can convolve the foreground and CMB PDFs to obtain a combined Gaussian PDF with variance $\sigma_{\rm fid}^2 = \sigma_{\rm CMB}^2 + \sigma_{\rm fg}^2$.  In the PDF analysis presented in Section~\ref{sec:interpretation}, we will marginalize over this variance, thus effectively desensitizing our analysis to any two-point information in the data.  We use $\sigma_{\rm fid}$ to set the fiducial level of this CMB+foreground variance.

This procedure accounts for the Gaussian contributions due to CIB, point sources, kSZ, and Galactic dust.  The lowest-order non-Gaussian contribution is the three-point function due to the CIB --- we test for any impact in our final results due to unmodeled non-Gaussian CIB contributions using the full CIB PDF measured from the simulations of~\cite{Sehgaletal2010}, and find no significant effects.  Higher-order non-Gaussian contributions exist due to the kSZ signal, which has zero skewness but non-zero kurtosis.  However, its amplitude is far smaller than the non-Gaussian moments of the tSZ signal~\cite{Hill-Sherwin2013} (and CIB), and its kurtosis is undetectable at ACT noise levels.  We test for its influence using the simulations of~\cite{Sehgaletal2010} and again find no detectable contamination.  Finally, gravitational lensing of the CMB in principle induces a non-zero kurtosis, but its amplitude is small enough to be undetectable in a full-sky, cosmic-variance limited experiment~\cite{Kesdenetal2002}.  We thus neglect it in our calculations.
\end{itemize}

\begin{figure}
\centering
\includegraphics[totalheight=0.5\textheight]{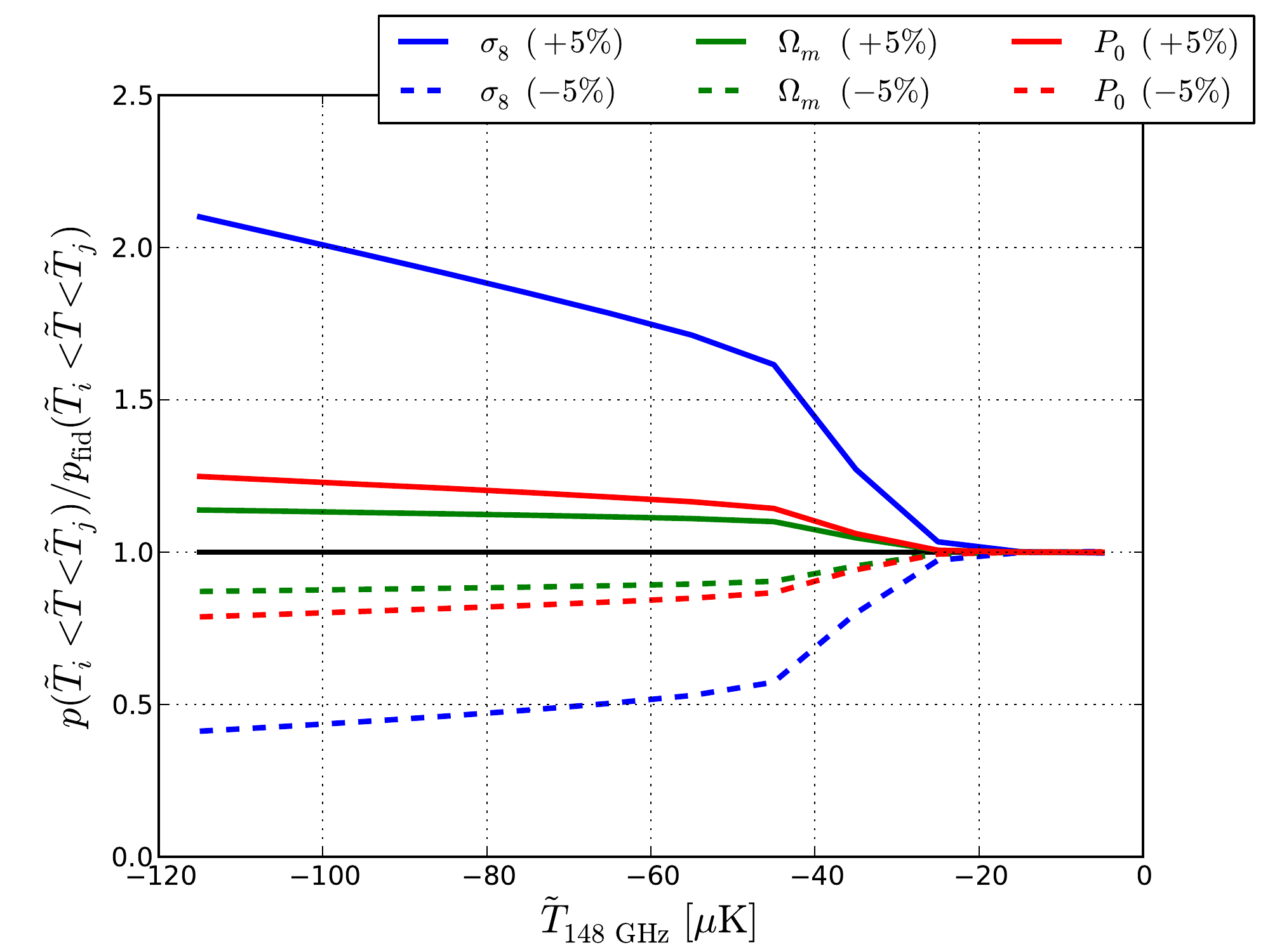}
\caption{Dependence of the tSZ PDF on cosmological and astrophysical parameters.  The plot shows the ratio with respect to the fiducial tSZ PDF (see Fig.~\ref{fig.PDFfid}) for $\pm 5$\% variations for each parameter: the amplitude of density perturbations $\sigma_8$, the matter density $\Omega_m$, and the dimensionless amplitude of the pressure--mass relation $P_0$.  The sensitivity to $\sigma_8$ is quite pronounced in the tSZ-dominated tail, with the values in these bins scaling as $\sigma_8^{10}$ to $\sigma_8^{16}$.  As expected, the bins in the noise-dominated regime are almost completely insensitive to variations in the cosmological or astrophysical parameters.
\label{fig.paramdep}}
\end{figure}

Combining these results, we construct the total noise kernel, $N(\tilde{T}-\tilde{T}')$, by convolving all of the components with one another:
\beq
N(\tilde{T}-\tilde{T}') = \int d\tilde{T}'' N_{\rm inst}(\tilde{T}-\tilde{T}'-\tilde{T}'') \int d\tilde{T}''' N_{\rm CMB}(\tilde{T}''-\tilde{T}''') N_{\rm fg}(\tilde{T}''') \,.
\label{eq.Ntot}
\eeq
This result for $N(\tilde{T}-\tilde{T}')$ can be directly applied to Eq.~(\ref{eq.noisekernel}) to compute the probability of observing a signal in bin $b_i$ given an input signal $\tilde{T}$, which completes the calculation of the noise-convolved, observable tSZ PDF in Eq.~(\ref{eq.yPDF}).

Fig.~\ref{fig.PDFfid} shows the observable tSZ PDF computed for our fiducial cosmology (WMAP9) and pressure profile ($P_0 = 1$, see Section~\ref{sec:theorymodel}), with the contributions from the tSZ signal (i.e., the second term in Eq.~(\ref{eq.yPDF})) and the noise (i.e., the first term in Eq.~(\ref{eq.yPDF})) shown individually.  The lowest-$|\tilde{T}|$ bins near the peak of the PDF are dominated by the noise, as expected, while a clear non-Gaussian tail sourced by the tSZ effect extends to negative $\tilde{T}$ values.  For reference, the figure shows the temperature decrement corresponding to a $5\sigma$ excursion in the ACT Equatorial map; the tSZ PDF includes contributions both above and below this point.  We also verify the claim made above that the total observable ACT PDF is insensitive to the lower mass cutoff used in the integrals of Eq.~(\ref{eq.yPDF}), whose fiducial value is $M_{\rm min} = 2 \times 10^{14} M_{\odot}/h$.  If we change the cutoff to $M_{\rm min} = 1.5 \times 10^{14} M_{\odot}/h$ ($M_{\rm min} = 3 \times 10^{14} M_{\odot}/h$), the mean of the ratio between the fiducial PDF in Fig.~\ref{fig.PDFfid} and the PDF in this test is 1.003 (0.993), after averaging over the twelve $\tilde{T}$ bins.  Slight changes to the lower mass cutoff simply shift the signal from low-mass clusters (whose signal is well below the noise in the map) from one term in Eq.~(\ref{eq.yPDF}) to the other.  This result indicates that our theoretical calculations are under control and properly converged.  


Fig.~\ref{fig.paramdep} presents the dependence of the noise-convolved, observable tSZ PDF on the most relevant cosmological and astrophysical parameters in our model, showing the ratio with respect to the fiducial case as each parameter is increased or decreased by $5$\%.  As is well-known, nearly all tSZ observables are very sensitive to $\sigma_8$~(e.g.,~\cite{Komatsu-Seljak2002,Hill-Sherwin2013,Bhattacharyaetal2012}), and this is true for the tSZ PDF as well.  Quoting simple power-law scalings, in the most negative $\tilde{T}$ bin shown in Fig.~\ref{fig.paramdep}, the PDF scales as $\sigma_8^{16}$, with the dependence remaining quite steep (from $\sigma_8^{10}$ to $\sigma_8^{15}$) in all bins until the noise contributions become important.  These scalings compare favorably to those for the tSZ power spectrum ($\sigma_8^{7-8}$)~\cite{Komatsu-Seljak2002,Tra11,Shawetal2010,Hill-Sherwin2013} or bispectrum ($\sigma_8^{10-12}$)~\cite{Hol07,Bhattacharyaetal2012,Hill-Sherwin2013}.

The signal also depends non-trivially on $\Omega_m$, the matter density, and $P_0$, the normalization of the pressure--mass relation as specified in Eq.~(\ref{eq.P0def}).  It is more sensitive to the latter, with the PDF in the most negative $\tilde{T}$ bin shown in Fig.~\ref{fig.paramdep} scaling as $P_0^{4.6}$, while the same bin scales with $\Omega_m$ as $\Omega_m^{2.5}$.  We find that $\sigma_8$ and $\Omega_m$ are the most relevant cosmological parameters.  For the ICM model, the PDF's dependence on $P_0$ is comparable to that on the mean outer logarithmic slope of the pressure profile (see~\cite{Hill-Pajer2013} for similar calculations).  Due to the computational cost and the SNR of our measurement, we allow only $\sigma_8$, $\Omega_m$, and $P_0$ to vary in the model fitting in Section~\ref{sec:interpretation}.  Future analyses with higher SNR may be able to simultaneously constrain multiple parameters.  A key point to note in Fig.~\ref{fig.paramdep} is that the ratio of the scaling with $\sigma_8$ and $P_0$ changes across the bins.  In other words, these parameters are not completely degenerate in the tSZ PDF.  This result suggests that a high-precision measurement of the tSZ PDF can simultaneously constrain cosmology ($\sigma_8$) and the pressure--mass relation ($P_0$), breaking the degeneracy between these quantities that is currently the limiting systematic in cluster cosmology analyses~(e.g.~\cite{Hill-Sherwin2013,Hasselfieldetal2013,Hill-Spergel2014,Planck2013counts,Reichardtetal2013}).  Unfortunately, the ACT data in our analysis are not quite at the level needed to strongly break the cosmology--ICM degeneracy, but future analyses should be able to do so.  Finally, although we have quoted simple scalings for the PDF values in some bins, we evaluate the tSZ PDF directly using Eq.~(\ref{eq.yPDF}) at each sampled point in parameter space.

\begin{figure}
\begin{minipage}[b]{0.495\linewidth}
\centering
\includegraphics[width=\textwidth]{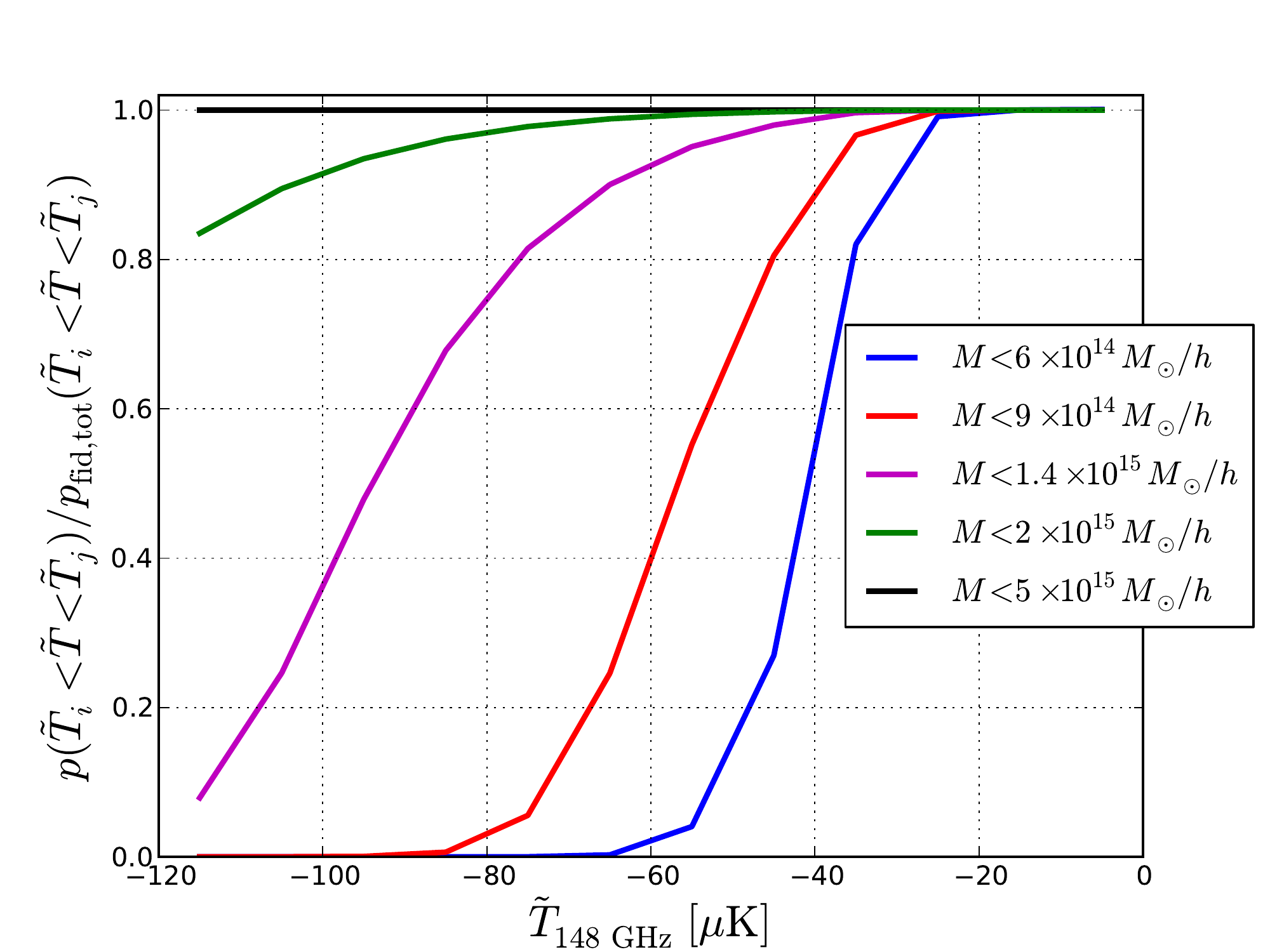}
\end{minipage}
\begin{minipage}[b]{0.495\linewidth}
\centering
\includegraphics[width=\textwidth]{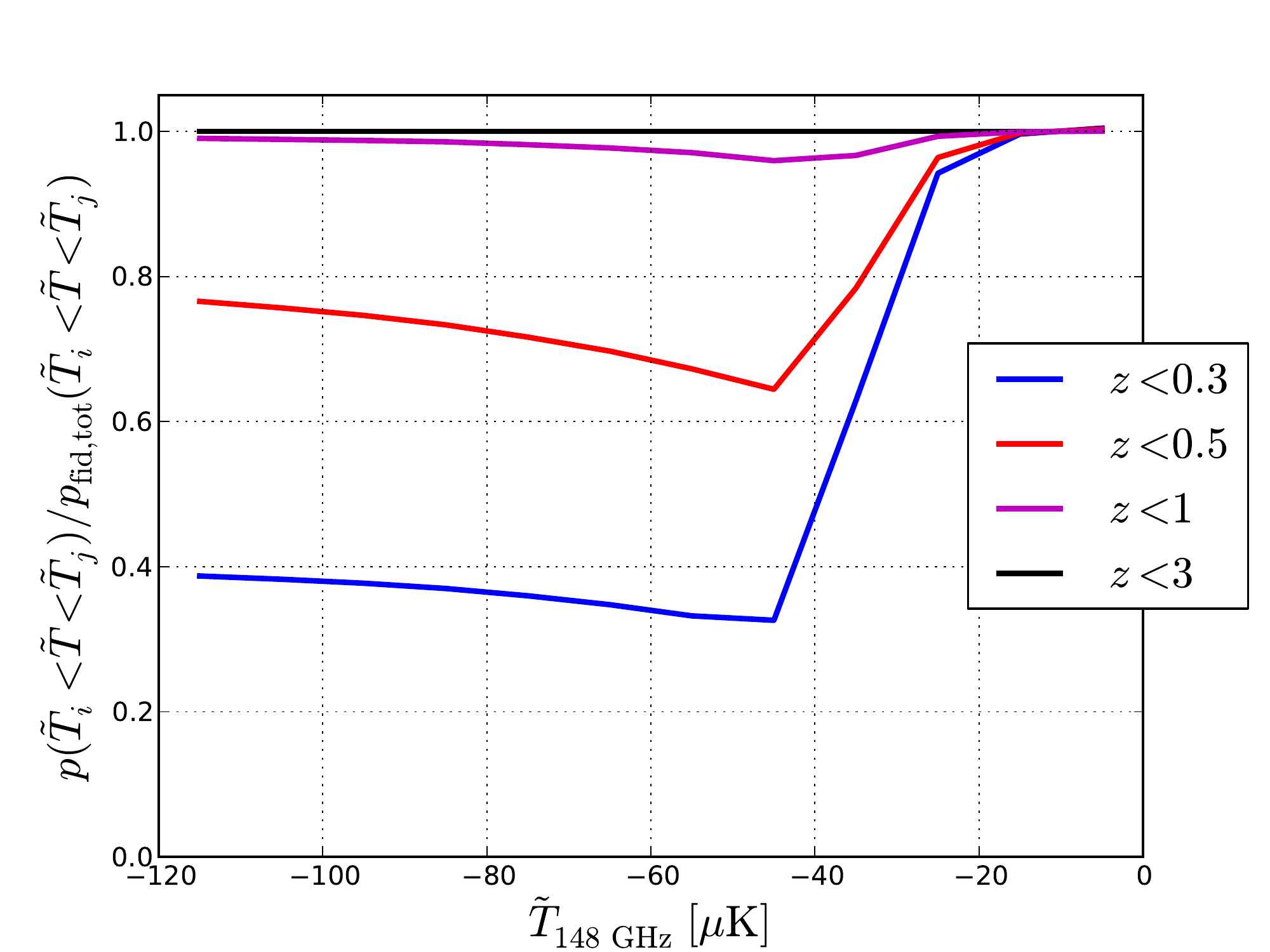}
\end{minipage}
\caption{Mass and redshift contributions to the noise-convolved, observable tSZ PDF.  The curves show the fraction of the total signal in each bin that is sourced by clusters at a mass or redshift below the specified value.  In the small-$|\tilde{T}|$ bins, the total signal is dominated by noise (see Fig.~\ref{fig.PDFfid}), and thus there is essentially no information about the mass or redshift contributions.  Below $\tilde{T} \approx -40 \, \mu$K, the tSZ signal dominates over the noise, and robust inferences can be made about the mass and redshift contributions, as shown.  These results are specific to the noise levels and angular resolution of the filtered ACT Equatorial map, and must be recomputed for different experimental scenarios.
\label{fig.masscontrib}}
\end{figure}

Fig.~\ref{fig.masscontrib} shows the contributions to the noise-convolved, observable tSZ PDF in our fiducial calculation from various mass and redshift ranges.  The noise-dominated bins at small $|\tilde{T}|$ values contain effectively no information about the mass and redshift contributions.  Proceeding into the tSZ-dominated tail, the plots demonstrate the characteristic mass and redshift scales contributing to the signal.  As expected, the outermost bins in the tail are sourced by the most massive clusters in the universe ($M \gtrsim 10^{15} \, M_{\odot}/h$), and hence are dominated by low-redshift contributions ($z \lsim 0.5$), since such clusters are rare at high redshift.  The first tSZ-dominated bins beyond the noise ($|\tilde{T}| \approx 40$--$50 \, \mu$K) are sourced by moderately massive clusters ($2 \times10^{14} \, M_{\odot}/h\lsim M \lsim 9 \times10^{14} \, M_{\odot}/h$), with $\approx 30$--$40$\% of the signal coming from $z \gtrsim 0.5$.  These statements are entirely dependent on the filter applied here, as well as the noise level and angular resolution of the experiment under consideration; an experiment with much lower noise and/or higher resolution than ACT would probe the PDF sourced by progressively lower-mass, higher-redshift objects.  The results here are specific to the ACT Equatorial 148 GHz map.

These results indicate that the theoretical modeling uncertainty for the objects dominating the tSZ PDF signal in the ACT map should not be overwhelming --- there are many observational constraints on massive, low-redshift clusters, and theoretical considerations indicate that the thermodynamics of such objects should be dominated by gravitation rather than poorly understood input from active galactic nuclei, turbulence, and other mechanisms~(e.g.,~\cite{Shawetal2010,Battagliaetal2012b}).  Nonetheless, a substantial fraction of the signal arises from objects below the threshold for direct detection in blind mm-wave cluster surveys: for the ACT Stripe 82 cluster sample, the approximate $90$\% completeness threshold for clusters detected at SNR~$> 5$ (for which optical confirmation is $100$\% complete for $z<1.4$) is $M_{500} \approx 5 \times 10^{14} \, M_{\odot}/h_{70} = 3.5 \times 10^{14} \, M_{\odot}/h$ over $0.2 \lsim z \lsim 1.0$ (low-redshift clusters are difficult to detect due to confusion with primordial CMB anisotropies).  Converting to the virial mass definition used here, this limit corresponds to roughly $M \approx 6$--$6.5 \times 10^{14} \, M_{\odot}/h$, depending on the redshift considered.  However, the incompleteness increases dramatically for $z<0.2$, so these numbers should only be taken as an approximate guide (see Section 3.6 and Fig.~11 of~\cite{Hasselfieldetal2013} for full details).  Any additional statistical constraining power for the tSZ PDF compared to the number counts arises from the inclusion of objects that are potentially missing in the latter, in particular $z \lsim 0.2$ groups and clusters over a wide range of masses or low- to moderate-mass objects over a wide redshift range.

\begin{figure}
\begin{minipage}[b]{0.495\linewidth}
\centering
\includegraphics[width=\textwidth]{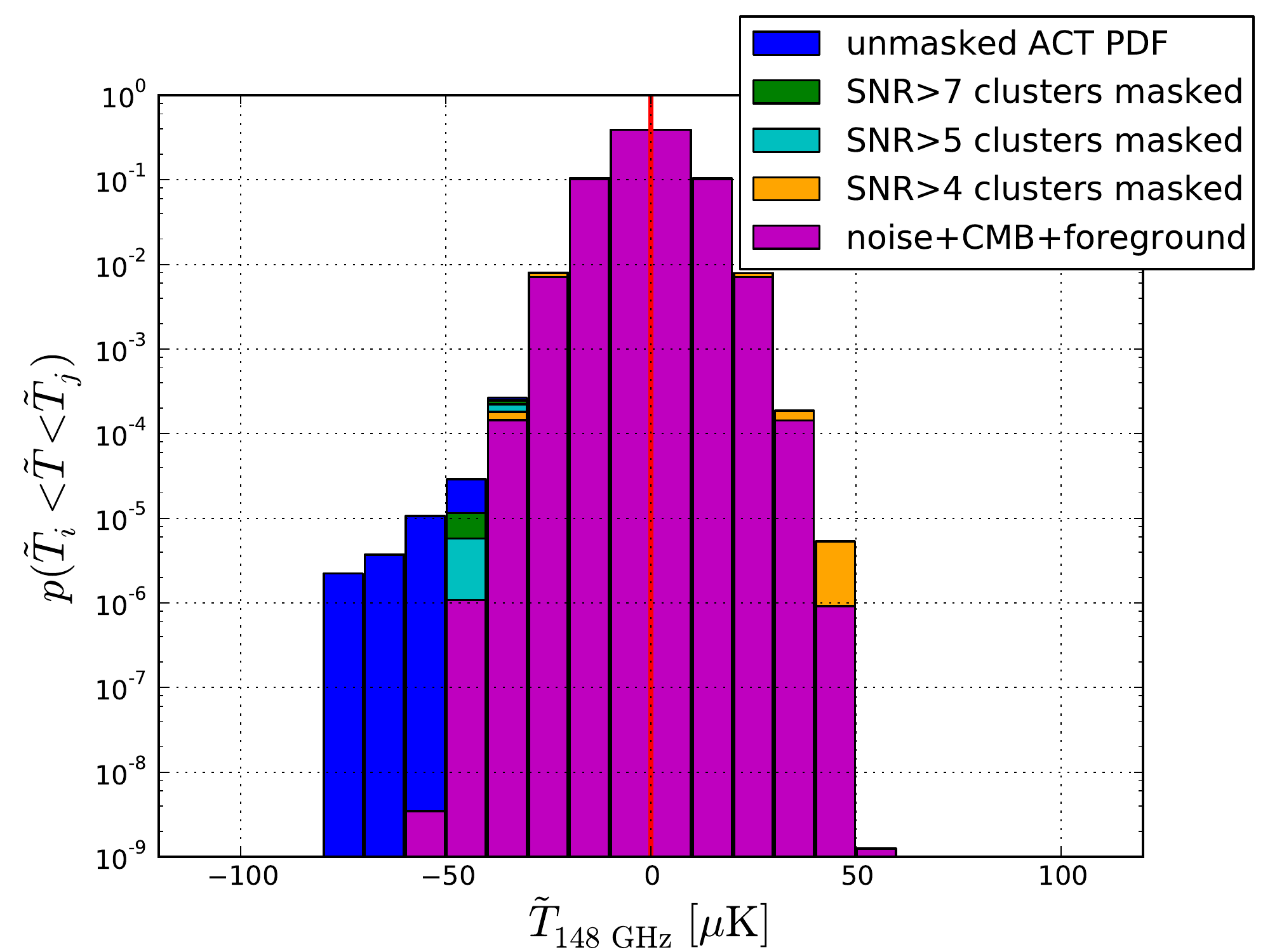}
\end{minipage}
\begin{minipage}[b]{0.495\linewidth}
\centering
\includegraphics[width=\textwidth]{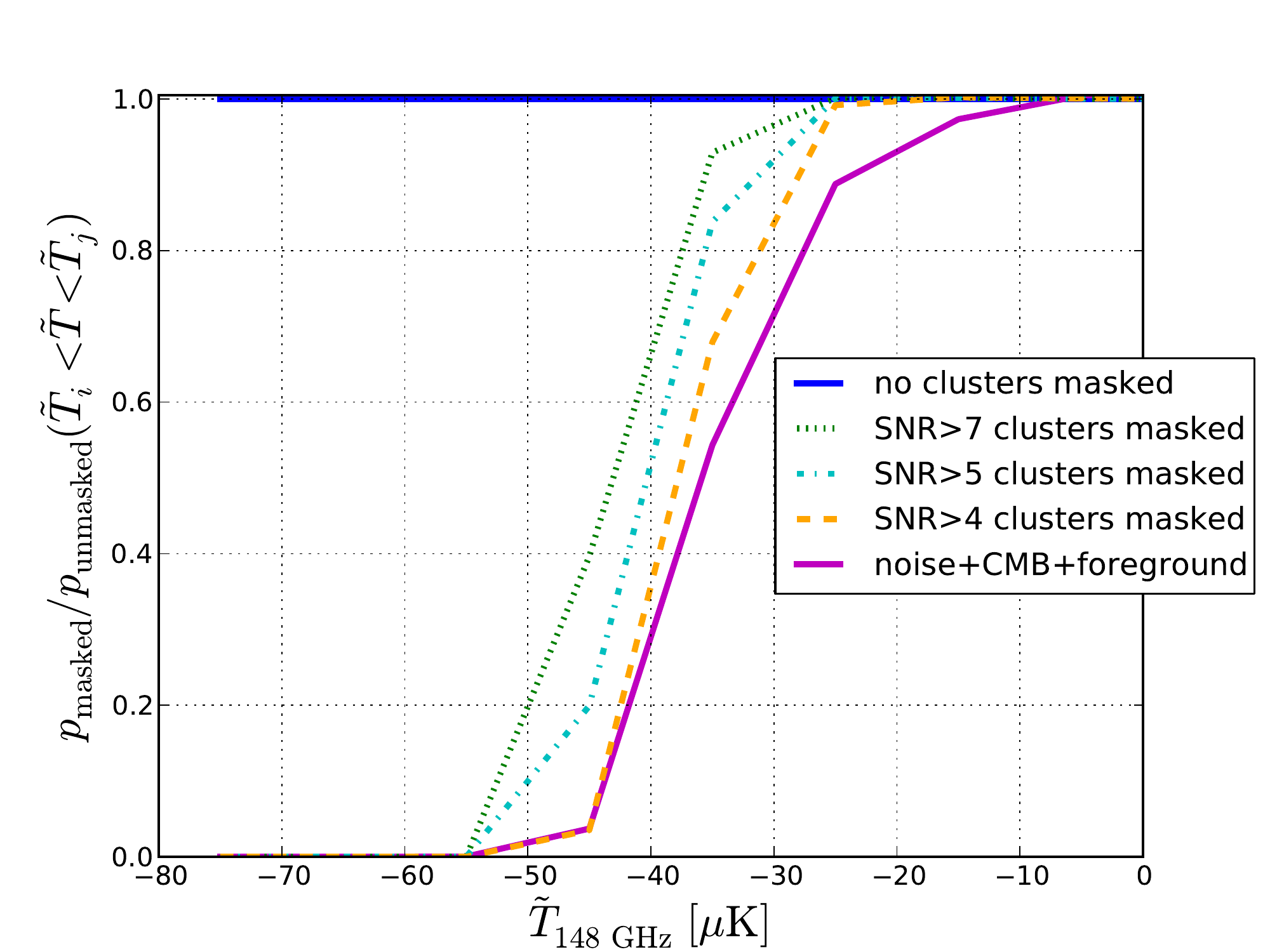}
\end{minipage}
\caption{Contributions to the observed ACT PDF from clusters of progressively lower SNR in the detected ACT Equatorial cluster catalog~\cite{Hasselfieldetal2013}, with SNR thresholds labeled in the figure.  The histograms in the left panel show the measured PDF as clusters above a given SNR threshold are masked.  For comparison, we also show the fiducial noise model in magenta, including instrumental and atmospheric noise, CMB, and non-tSZ foregrounds.  The right panel shows the same information in terms of a ratio with respect to the original, unmasked ACT PDF, but showing only the $\tilde{T}<0$ region of interest.  As anticipated, the largest decrements in the PDF are due to the highest SNR clusters in the detected catalog, but clear contributions above the noise model also arise from objects below the threshold for direct detection (SNR~$<5$).
\label{fig.maskHass}}
\end{figure}

To further investigate these points, Fig.~\ref{fig.maskHass} shows how the observed ACT PDF changes as clusters from the catalog presented in~\cite{Hasselfieldetal2013} are masked to a progressively lower SNR threshold.  Specifically, the SNR in this exercise is the SNR for each cluster in a matched-filtered map constructed using the $\theta_{500} = 5.9'$ filter in~\cite{Hasselfieldetal2013}, rather than the maximum SNR for each cluster found over all considered filter scales.  The fixed filter scale SNR allows a more robust definition over the full cluster sample.  For a given SNR threshold, we locate all clusters in the ACT Equatorial map above this cut and then mask all pixels in the map within an angular distance of $5\theta_{500} = 5 r_{500}/d_A(z)$ from the center of each cluster.  The radius $r_{500}$ is computed from the mass $M_{500}$ determined using the ``universal pressure profile'' scaling relation (see Section 3.2 and Table 8 of~\cite{Hasselfieldetal2013}).  The left panel of Fig.~\ref{fig.maskHass} shows the measured ACT PDF as clusters are masked in this fashion down to a SNR threshold of 7, 5, and 4.  In addition, we show the fiducial noise model, including instrumental and atmospheric noise, CMB, and non-tSZ foregrounds, as described above.  The right panel of Fig.~\ref{fig.maskHass} shows the same information, but presented as a ratio with respect to the original, unmasked ACT PDF.  As expected, the largest decrements in the PDF are due to the highest SNR clusters detected in the Equatorial map.  However, the PDF also clearly contains signal from objects below the threshold for direct detection (compare the SNR~$>5$ and SNR~$>4$ masking thresholds to the noise model).  The contributions from these lower mass objects are notable in the $-50\, \mu {\rm K} < \tilde{T} < -25\, \mu {\rm K}$ range.  The part of the tSZ signal captured by the PDF that is missed in cluster count analyses is the difference between the measured PDF with SNR~$>5$ clusters masked (cyan) and the noise PDF (magenta), which can be clearly seen in the right panel of Fig.~\ref{fig.maskHass}.  This corresponds to thousands of pixels in the ACT map, some of which are actual low-mass clusters, and some of which are noise.  The presence of these low-mass objects can be inferred statistically in the PDF. 

\subsection{Covariance Matrix}
\label{sec:covmat}
In the limit of uncorrelated, homogeneous noise, the covariance matrix of the observable tSZ PDF can be computed using the analytic halo model framework described in the previous section.  The covariance matrix receives contributions from the Poisson statistics of the finite map pixelization, the Poisson statistics of the clusters, and the cosmic variance of the underlying density realization (often called the ``halo sample variance'' (HSV) term in other contexts)~\cite{Hu-Kravtsov2003,Takada-Bridle2007,Takada-Spergel2013}.  The HSV term is more relevant for small $|\tilde{T}|$-values, where less massive clusters become important in the PDF; this is analogous to the situation for the tSZ power spectrum covariance matrix, where the HSV term becomes relevant at high-$\ell$ where less massive halos dominate the power spectrum (see Fig.~1 in~\cite{Zhang-Sheth2007}).  The pixel Poisson term contributes only to the diagonal elements of the covariance matrix.  The cluster Poisson term, which arises from the finite sampling of the density field, dominates in the large-$|\tilde{T}|$ tail of the PDF.  The off-diagonal components of the tSZ PDF covariance matrix are non-trivial, as a single cluster contributing to multiple bins in the PDF produces obvious bin-to-bin correlations.  Using the notation defined in the previous section, the covariance matrix of the tSZ PDF, ${\rm Cov}_{ij} \equiv \langle p_i p_{j} \rangle - \langle p_i \rangle \langle p_{j} \rangle$, is given by:
\beqn
\langle p_i p_{j} \rangle - \langle p_i \rangle \langle p_{j} \rangle & = & \frac{\langle p_i \rangle}{N_{\rm pix}} \delta_{ij} \, + \hspace{140pt} \mathrm{(pixel \,\, Poisson \,\, term)} \nonumber \\
  & & \frac{1}{4\pi f_{\rm sky}} \int dz \frac{d^2 V}{dz d\Omega} \int dM \frac{dn}{dM} \left( g_i(M,z) - \rho_i(0) \pi \theta_{out}^2(M,z) \right) \times  \nonumber \\
  & &\left( g_{j}(M,z) - \rho_{j}(0) \pi \theta_{out}^2(M,z) \right) + \hspace{40pt} \mathrm{(cluster \,\, Poisson \,\, term)} \nonumber \\
  & & \int dz \frac{d^2 V}{dz d\Omega} \left[ \int dM_1 \frac{dn}{dM_1} b(M_1,z) \left( g_i(M_1,z) - \rho_i(0) \pi \theta_{out}^2(M_1,z) \right) \right] \times \nonumber \\
  & & \left[ \int dM_2 \frac{dn}{dM_2} b(M_2,z) \left( g_{j}(M_2,z) - \rho_{j}(0) \pi \theta_{out}^2(M_2,z) \right) \right] \times \nonumber \\
  & & \int \frac{\ell d\ell}{2\pi} P_{\rm lin} \left(\frac{\ell}{\chi},z \right) \left| \tilde{W}(\ell \Theta_s) \right|^2 \,, \hspace{55pt} \mathrm{(cluster \,\, HSV \,\, term)}
\label{eq.yPDFCov}
\eeqn
where $N_{\rm pix} \approx 4.04 \times 10^6$ is the number of pixels in the map, $f_{\rm sky} \approx 0.00679$ is the total observed sky fraction, $b(M,z)$ is the linear halo bias, $P_{\rm lin}(k,z)$ is the linear matter power spectrum, and $\tilde{W}(\vec{\ell})$ is the Fourier transform of the survey window function, which is defined such that $\int d^2\theta W(\vec{\theta}) = 1$.  We have used the Limber approximation~\cite{Limber1954} to obtain the HSV term and have assumed a circular survey geometry for simplicity in the HSV term, with $4\pi f_{\rm sky} = \pi \Theta_s^2$.  We use the bias prescription of~\cite{Tinkeretal2010} and compute the linear matter power spectrum using CAMB.

The covariance matrix in Eq.~(\ref{eq.yPDFCov}) depends on the cosmological and astrophysical parameters in our model.  The correct approach would be to compute the PDF covariance matrix simultaneously with the PDF signal at each point in parameter space~(e.g.,~\cite{Eifleretal2009}).  However, due to the computational expense of this method (which is extreme when simulations are required --- see the next section), we choose to make the simplification that the covariance matrix depends only on $\sigma_8$.  Based on the results shown in Fig.~\ref{fig.paramdep} for the PDF signal, this approximation should be quite accurate, since the dependence of the covariance matrix on these parameters should scale roughly as the square of the PDF signal's dependence.  Indeed, the diagonal covariance matrix elements scale very steeply with $\sigma_8$ (from $\sigma_8^{10}$ to $\sigma_8^{38}$) in the tSZ-dominated bins in the tail of the PDF.  

The current analysis neglects the HSV term based on calculations using our fiducial model that indicate it contributes $\lsim 10$\% to the total variance in the bins over which the observed ACT PDF is sensitive to the tSZ signal.  The HSV term becomes more important when probing to lower masses, and in future analyses it should likely be included.  

In addition, it is possible that the SNR of the tSZ PDF could be improved by masking some massive, low-redshift clusters, which contribute significant sample variance to the total error (through the cluster Poisson term in Eq.~(\ref{eq.yPDFCov})).  Similar effects arise in the covariance matrix of the tSZ power spectrum~\cite{Shawetal2009,Hill-Pajer2013} and bispectrum~\cite{Bhattacharyaetal2012,Crawfordetal2014}.  However, the masking of clusters also introduces additional uncertainty into the measurement of tSZ statistics, as it requires knowledge of the $Y$--$M$ relation (i.e., pressure--mass relation), leading to an uncertainty in the masking threshold.  Masking massive clusters would also discard the PDF signal in the highest-$|\tilde{T}|$ bins, where the cosmological sensitivity is greatest.  Finding the optimal masking scenario requires balancing these competing considerations (see~\cite{Crawfordetal2014}), which we defer to future work.  We leave all tSZ signal unmasked in the ACT map here.

While it is useful to have a analytic prescription for the covariance matrix, in practice the underlying approximation of uncorrelated noise is not valid for our analysis of the ACT 148 GHz PDF.  Even if there were no instrumental or atmospheric noise, the CMB itself is a source of noise in our analysis, and it obviously has a non-trivial angular correlation function.  The $\ell$-space filtering applied to the maps in order to upweight the tSZ signal also has the byproduct of correlating the noise across pixels in real space.  Thus, realistic Monte Carlo simulations of the ACT data are required to estimate the full covariance matrix accurately.

\section{Simulations}
\label{sec:sims}
We construct realistic simulations of our data set to account for the effects of correlated, inhomogeneous noise in the covariance matrix and to estimate any biases in our likelihood analysis presented in Section~\ref{sec:interpretation}.  Although knowledge of the one-point PDF of the component signals (regardless of their spatial correlations) suffices to compute the total one-point PDF of the filtered ACT 148 GHz map, as described in Section~\ref{sec:noisyPDF}, this knowledge is not sufficient to compute the covariance matrix of the PDF in the presence of correlated noise.  It is straightforward to show that in this case estimation of the covariance matrix requires knowledge of the contribution of each pair of pixels in the map to the covariance between bin $i$ and bin $j$ (for each component signal in the map).  Assuming homogeneity, this would depend only on the distance between the two pixels for each pair; however, the noise in our map is not perfectly homogeneous.  We take the approach of simply simulating all relevant components in the map.  Computing the one-point PDF itself depends only on the average properties of a single pixel in the map --- no sum over pairs of pixels is required --- and hence a simple convolution of the component PDFs suffices to compute the total PDF, as outlined in Section~\ref{sec:noisyPDF}.  The computation of the covariance matrix in the presence of correlated, inhomogeneous noise necessitates simulations, which also allow tests of some assumptions in the analytic calculations of the PDF, including the approximation of no overlapping clusters along the line-of-sight.

The simulation pipeline is comprised of tools produced for earlier ACT studies, in particular the CMB temperature power spectrum and CMB lensing analyses~\cite{Dasetal2011b,Dasetal2011}, as well as new tSZ tools constructed specifically for this analysis.  The steps in the pipeline are as follows:
\begin{enumerate}
\item Generate a random CMB temperature field seeded by the angular power spectrum of our fiducial WMAP9 cosmology; this temperature field is then gravitationally lensed (which is not necessary for this analysis as argued in Section~\ref{sec:noisyPDF}, but was already built into the pipeline).  The details of these steps can be found in~\cite{Dasetal2011b}.
\item Generate a Gaussian random field seeded by the best-fit power spectrum of all non-tSZ foreground components as determined in~\cite{Sieversetal2013}, following the same method with which the random CMB temperature field is generated.  These foreground components are described in Section~\ref{sec:noisyPDF}.
\item Generate a map of Poisson-distributed tSZ clusters drawn from the Tinker mass function~\cite{Tinkeretal2008} integrated over the same mass and redshift limits used in our analytic calculations.  Compute the $T$-profile of these clusters (including relativistic corrections) using our fiducial pressure profile model from~\cite{Battagliaetal2012}, which is discretized on a grid of ACT-sized pixels (roughly $0.25 \, {\rm arcmin}^2$).
\item Sum the lensed CMB map, foreground map, and tSZ cluster map.
\item Convolve the summed map with the appropriate ACT 148 GHz beam~\cite{Hasselfieldetal2013beam} and add in randomly-seeded realistic ACT noise using the method described in Section 4 of~\cite{Dasetal2011b}.  The seed for the noise is generated from difference maps of subsets of the ACT Equatorial data and is scaled according to the number of observations in each region of the map, and hence accounts for all non-trivial correlation properties and inhomogeneities in the noise.  
\item Filter and process the final map using exactly the same procedure as used on the ACT 148 GHz data map (see Section~\ref{sec:data}).
\end{enumerate}

\begin{figure}
\centering
\includegraphics[totalheight=0.5\textheight]{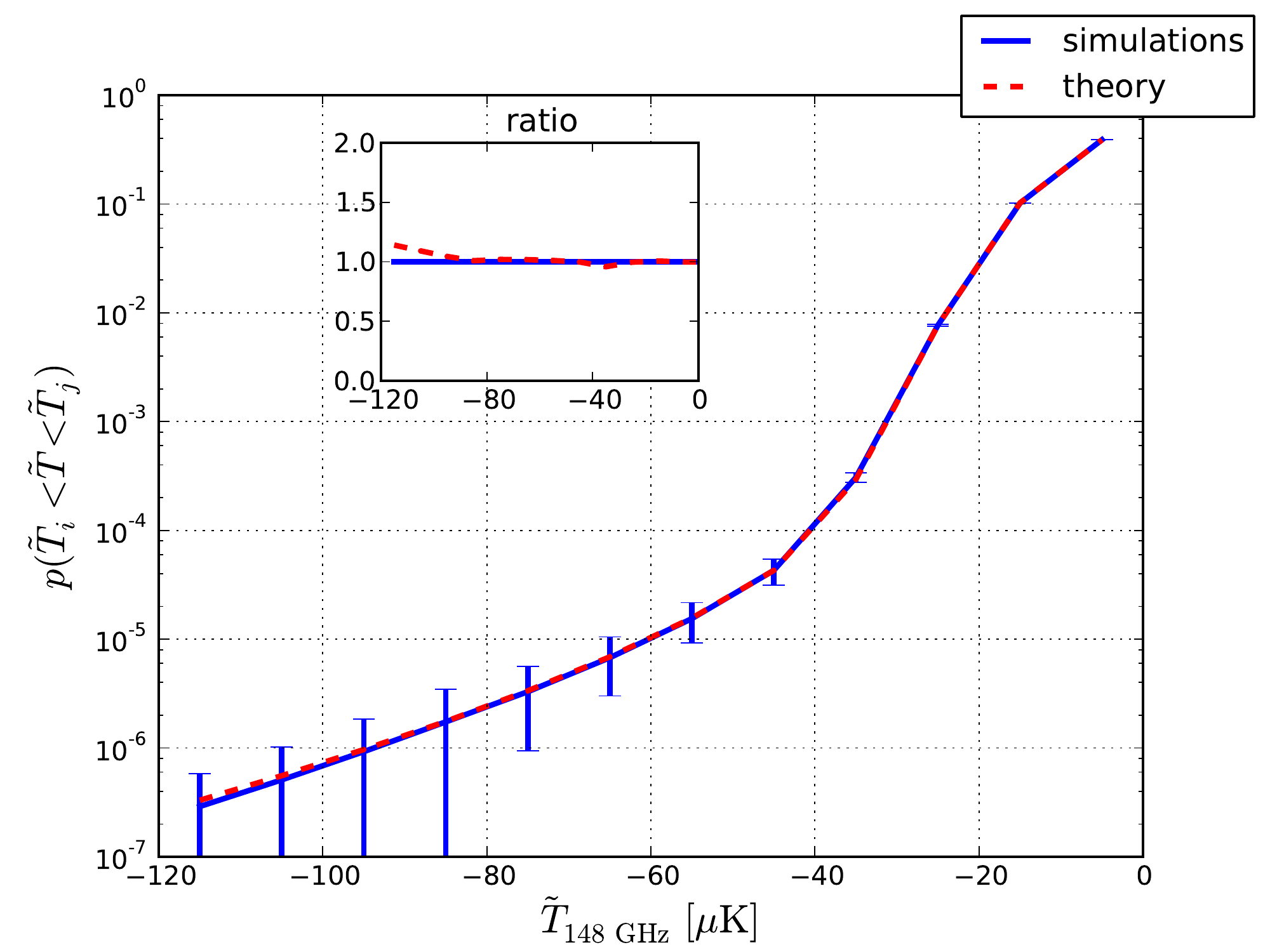}
\caption{Comparison of the analytic halo model computation of the noise-convolved, observable tSZ PDF using Eq.~(\ref{eq.yPDF}) (red dashed curve) to the PDF directly measured from the simulated maps described in Section~\ref{sec:sims} (blue solid curve).  The error bars are computed from the scatter amongst the 476 simulated maps.  The two curves are nearly indistinguishable --- the inset plot shows the ratio of the theory curve to the simulation curve.  This validates the halo model approximation that clusters do not overlap along the line-of-sight~(see Section~\ref{sec:tSZonlyPDF}), within the mass and redshift limits used in our analysis.  No free parameter is adjusted or fit in either calculation to force the simulation and theory curves shown here to agree.
\label{fig.simcomp}}
\end{figure}

\begin{figure}
\centering
\includegraphics[totalheight=0.5\textheight]{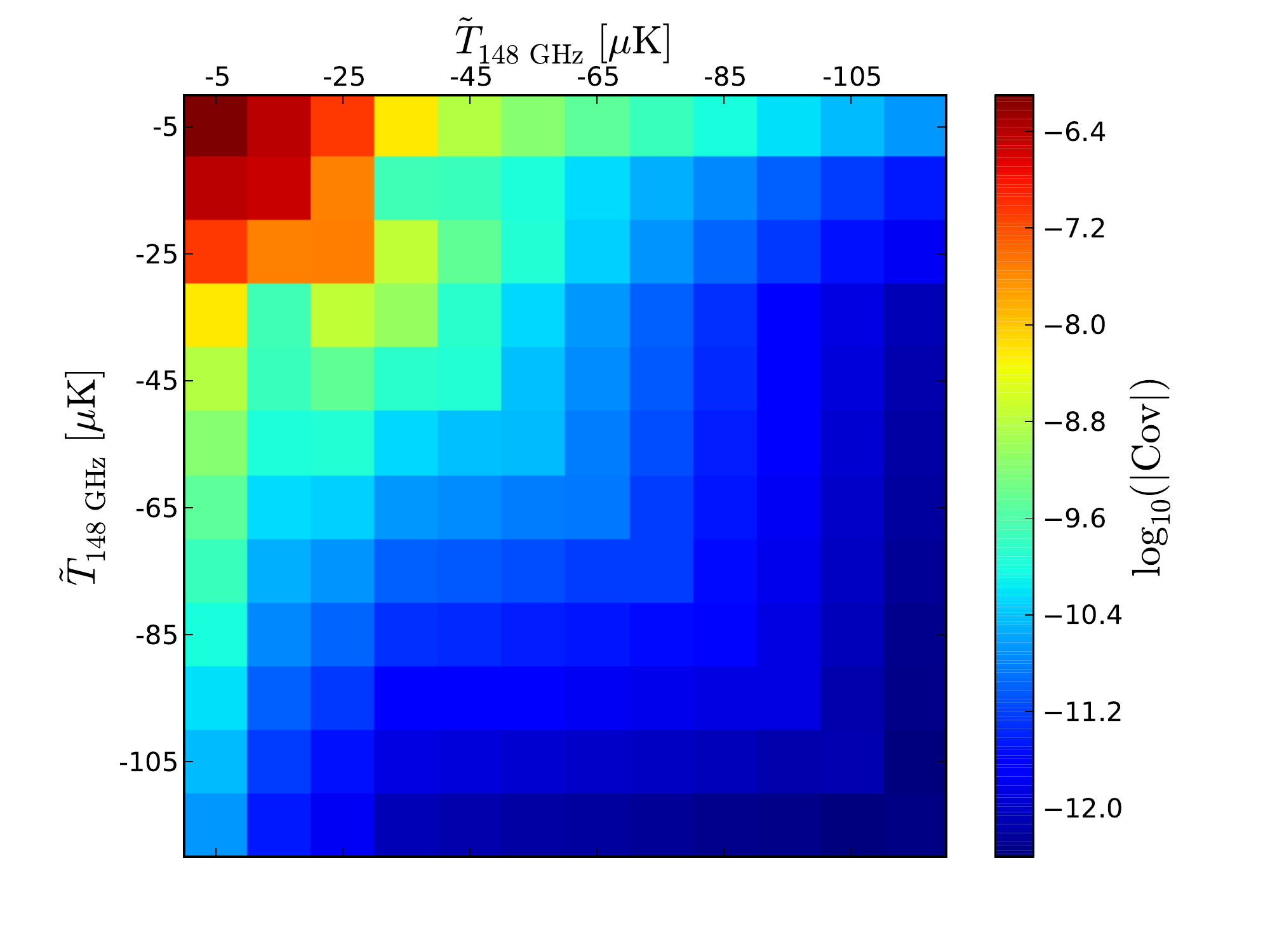}
\caption{Covariance matrix of the noise-convolved, observable tSZ PDF computed for our fiducial cosmology and pressure profile model using the analytic halo model framework described in Section~\ref{sec:covmat} with corrections computed from the Monte Carlo simulations described in Section~\ref{sec:sims}.  The quantity plotted is $\log_{10} \left( | {\rm Cov} | \right)$, where the absolute value is necessary due to negative values in some of the off-diagonal bins representing correlations between the noise-dominated region and the tSZ-dominated region of the PDF.  A significant amount of bin-to-bin correlation is evident, especially among the tSZ-dominated bins in the tail of the PDF.
\label{fig.PDFcovfid}}
\end{figure}

We generate 476 simulated ACT Equatorial maps using this procedure.  We then compute the mean PDF and the PDF covariance matrix using the 476 simulated maps.  Furthermore, since the covariance matrix of the PDF is a function of $\sigma_8$ (see the previous section), we repeat this process for $20$ values of $\sigma_8$ linearly spaced between $0.735$ and $0.858$.  The simulations account for all relevant signals in the $\tilde{T} < 0$ region of the 148 GHz PDF, with the possible exception of higher-order moments (3-point and higher) of the unresolved point source, CIB, or kSZ fields.

We verify that the tSZ cluster map is insensitive to the details of the pixelization by increasing the resolution by a factor of two (i.e., factor of four in area) and re-running the third step in the pipeline listed above.  We also verify that edge effects do not introduce any bias in the results.  The simulated maps allow us to directly quantify the effect of overlapping clusters along the line-of-sight: for our fiducial mass and redshift limits, only about $10$\% of the sky area populated by clusters (which is itself about $30$\% of the full sky, i.e., $F_{\rm clust} \approx 0.3$ for $2 \times 10^{14} M_{\odot}/h < M < 5 \times 10^{15} M_{\odot}/h$) consists of line-of-sight overlaps of multiple objects.  This test verifies that the assumptions in our analytic calculations are valid.  We demonstrate this consistency directly in Fig.~\ref{fig.simcomp}, which displays the tSZ PDF for our fiducial cosmology and pressure profile computed using the analytic method in Section~\ref{sec:noisyPDF} and using the simulations described above.  The results are in excellent agreement.

In order to obtain the covariance matrix of the tSZ PDF for use in the likelihood analysis in the next section, we combine the smooth analytic results found using Eq.~(\ref{eq.yPDFCov}) with the covariance matrices obtained from the simulations.  Attempts to use the simulated covariance matrices directly lead to ill-behaved likelihood functions due a lack of convergence in the far off-diagonal elements of the covariance matrix.  Although 476 simulations should suffice to determine the covariance matrix of 12 (correlated) Gaussian random variables, the values of the PDF in the tSZ-dominated bins are not perfectly Gaussian-distributed, especially in the largest-$|\tilde{T}|$ bins, which are sourced by rare, massive clusters.  Moreover, the convergence of the simulation-estimated covariance matrix depends on the value of $\sigma_8$, since many fewer clusters are distributed for low $\sigma_8$ values compared to high $\sigma_8$ values, thus requiring many more simulations to achieve convergence.

Motivated by these issues, in practice we combine the smooth analytic covariance matrices computed using Eq.~(\ref{eq.yPDFCov}) with corrections computed from the simulations.  The corrections are small in the tSZ-dominated tail of the PDF ($\approx 5-20$\%), but can be as large as a factor of $\approx 10$ in the noise-dominated bins at small $|\tilde{T}|$.  
In particular, the diagonal elements of the simulated covariance matrices are well-converged for all values of $\sigma_8$, as are all of the small-$|\tilde{T}|$ elements (on- and off-diagonal), since those bins are highly populated in all realizations.Ê For the inner three bins, we fit the ratio of simulated to analytic covariance matrices as a linear function of $\sigma_8$. For the outer nine bins, we fit only the diagonal elements; for the off-diagonal elements linking the inner and outer sub-matrices, we apply a correction given by the geometric mean of the two corresponding diagonal elements.  We then apply the correction factors computed using these linear fits to each of the analytic covariance matrices as a function of $\sigma_8$.  The final result is an estimate of the covariance matrix that is a smooth function of $\sigma_8$ and simultaneously includes the effects of correlated, inhomogeneous noise captured in the simulations.  Fig.~\ref{fig.PDFcovfid} presents the covariance matrix computed for our fiducial cosmology and pressure profile.  In particular, bin-to-bin correlations induced by correlated noise and by clusters that contribute tSZ signal to multiple bins are clearly visible.

\section{Interpretation}
\label{sec:interpretation}
\subsection{Likelihood Function}
\label{sec:like}
We use the measurement of the tSZ PDF in the ACT Equatorial 148 GHz data to constrain cosmological parameters.  Given that $\sigma_8$ is the most relevant cosmological parameter by a significant margin (see Fig.~\ref{fig.paramdep}), we start by allowing it alone to vary amongst the $\Lambda$CDM parameters.  We will further consider scenarios in which the overall normalization of the pressure--mass relation, $P_0$, is either fixed or is free to vary, and eventually we will free $\Omega_m$ as well.  Finally, we marginalize all results over a nuisance parameter corresponding to the variance in the PDF, $\sigma^2_{\rm nuis}$.  The fiducial value of this nuisance parameter is $\sigma_{\rm nuis} = \sigma_{\rm fid}$ given in Section~\ref{sec:noisyPDF}.  By marginalizing over it, we desensitize our results to any unknown Gaussian component in the map, including the contribution of any residual non-tSZ foregrounds or the effect of a primordial CMB variance differing slightly from our fiducial WMAP9 assumption.  We place a Gaussian prior of width $0.1 \, \mu {\rm K}$ on $\sigma_{\rm nuis}$, since its fiducial value is computed using the CMB power spectrum and the best-fit non-tSZ foreground power spectra from~\cite{Sieversetal2013}, which are directly measured in the Equatorial map.  We verify that the maximum-likelihood values of $\sigma_{\rm nuis}$ are generally close to the fiducial value.  The primary motivation for $\sigma_{\rm nuis}$ is to capture any small amount of residual CIB or kSZ that could leak into the $\tilde{T} < 0$ PDF.  In order to further prevent this leakage, we discard the smallest $|\tilde{T}|$ bin in the PDF in the likelihood analysis, which corresponds to roughly the $1\sigma$ fluctuations in the map (the rms of the filtered map is $8.28 \, \mu$K, and the bins are $10 \, \mu$K wide).  The simulations of~\cite{Sehgaletal2010} suggest that the signal from non-tSZ foregrounds should generally all lie within this bin.  We are thus left with 11 bins, most of which are dominated by tSZ signal.

The likelihood function is
\beq
\mathcal{L}(\vec{\Theta}) = \frac{1}{\sqrt{(2\pi)^{N_b} \mathrm{det}(\mathrm{Cov}(\sigma_8))}} \mathrm{exp}\left(-\frac{1}{2} \left( p_i(\vec{\Theta}) - \hat{p}_i \right) \left(\mathrm{Cov}^{-1}(\sigma_8) \right)_{ij} \left( p_j(\vec{\Theta}) - \hat{p}_j \right) \right) \,,
\label{eq.like}
\eeq
where $N_b$ is the number of bins in the measurement, ${\rm Cov}_{ij}$ is the covariance matrix described in the previous section (computed as a function of $\sigma_8$ only), and $p_i$ is the PDF value in bin $i$, computed as a function of parameters $\vec{\Theta} = \{\sigma_8,\Omega_m,P_0,\sigma^2_{\rm nuis}\}$ (note that we will consider cases where some of these parameters are held fixed).  The likelihood function in Eq.~(\ref{eq.like}) relies on the assumption of Gaussianity, which breaks down in the bins far in the tSZ-dominated tail of the PDF, which are rarely populated (in fact, the outermost four bins in the ACT PDF are empty).  An approach based on Poisson statistics may be more appropriate for these bins, but this raises questions about properly accounting for bin-to-bin correlations, which Fig.~\ref{fig.PDFcovfid} indicates are strong.  The only reference we are aware of in the literature that presents a likelihood for the PDF in general (not the tSZ PDF in particular) is~\cite{Huffenberger-Seljak2005}.  However, their likelihood requires that the pixels --- and hence bins --- be uncorrelated, and so they re-scale the likelihood, effectively averaging over patches larger than the correlation length in the filtered map.  This compromise reduces the potential precision of the measurement.  In the absence of a more suitable likelihood, we use Eq.~(\ref{eq.like}) and search for any resulting bias in our constraints using the Monte Carlo simulations described in Section~\ref{sec:sims}.

Unfortunately, we do find a bias in constraints on $\sigma_8$ and $P_0$ when na\"{i}vely implementing Eq.~(\ref{eq.like}), especially in the case when both parameters are allowed to vary.  The bias pushes the recovered values of $\sigma_8$ ($P_0$) to lower (higher) results than the input values specified in the Monte Carlo simulations.  To verify that this bias is caused by the breakdown of the assumption of Gaussianity in the likelihood function, we ``Gaussianize'' the analysis by combining bins in the far tSZ-dominated tail of the PDF, creating a single, well-populated bin that obeys Gaussian statistics rather than Poisson statistics.  We progressively combine bins in the tail of the PDF until the bias is no longer present, eventually finding that combining the outermost 6 bins is sufficient.  Thus, in the likelihood in Eq.~(\ref{eq.like}), we have $N_b = 6$, where the sixth bin spans $[-120\, \mu {\rm K}, -60\, \mu {\rm K}]$ (recall that the first bin spans $[-20\, \mu {\rm K}, -10\, \mu {\rm K}]$, as we discard the bin nearest the PDF peak to obviate any leakage of CIB or point source emission from the positive side to the negative side of the PDF).  We apply the appropriate linear transformation to modify the covariance matrices computed in Section~\ref{sec:sims} to account for the final binning choice.  As an unfortunate byproduct of this need to ``Gaussianize'' the likelihood, the power of the ACT PDF to simultaneously constrain $\sigma_8$ and $P_0$ is substantially weakened, simply because the shape of the PDF is not as well constrained when combining so many smaller bins into a single larger bin.  A clear goal for future PDF analyses is to implement a more sophisticated, non-Gaussian likelihood function, allowing the full use of the constraining power in the PDF.

\subsection{Constraints}
We apply the likelihood described in the previous section to the pixel temperature histogram of the filtered, processed ACT Equatorial 148 GHz maps described in Section~\ref{sec:data}.  
In all constraints quoted in the following, we report the best-fit value as the mean of the marginalized likelihood, while the lower and upper error bounds correspond to the $16\%$ and $84\%$ points in the marginalized cumulative distribution, respectively.  The considered parameter ranges for $\sigma_8$, $\Omega_m$, and $P_0$ are $[0.670,0.885]$, $[0.212,0.362]$, and $[0.5,1.5]$, respectively.

\begin{figure}
\centering
\includegraphics[totalheight=0.5\textheight]{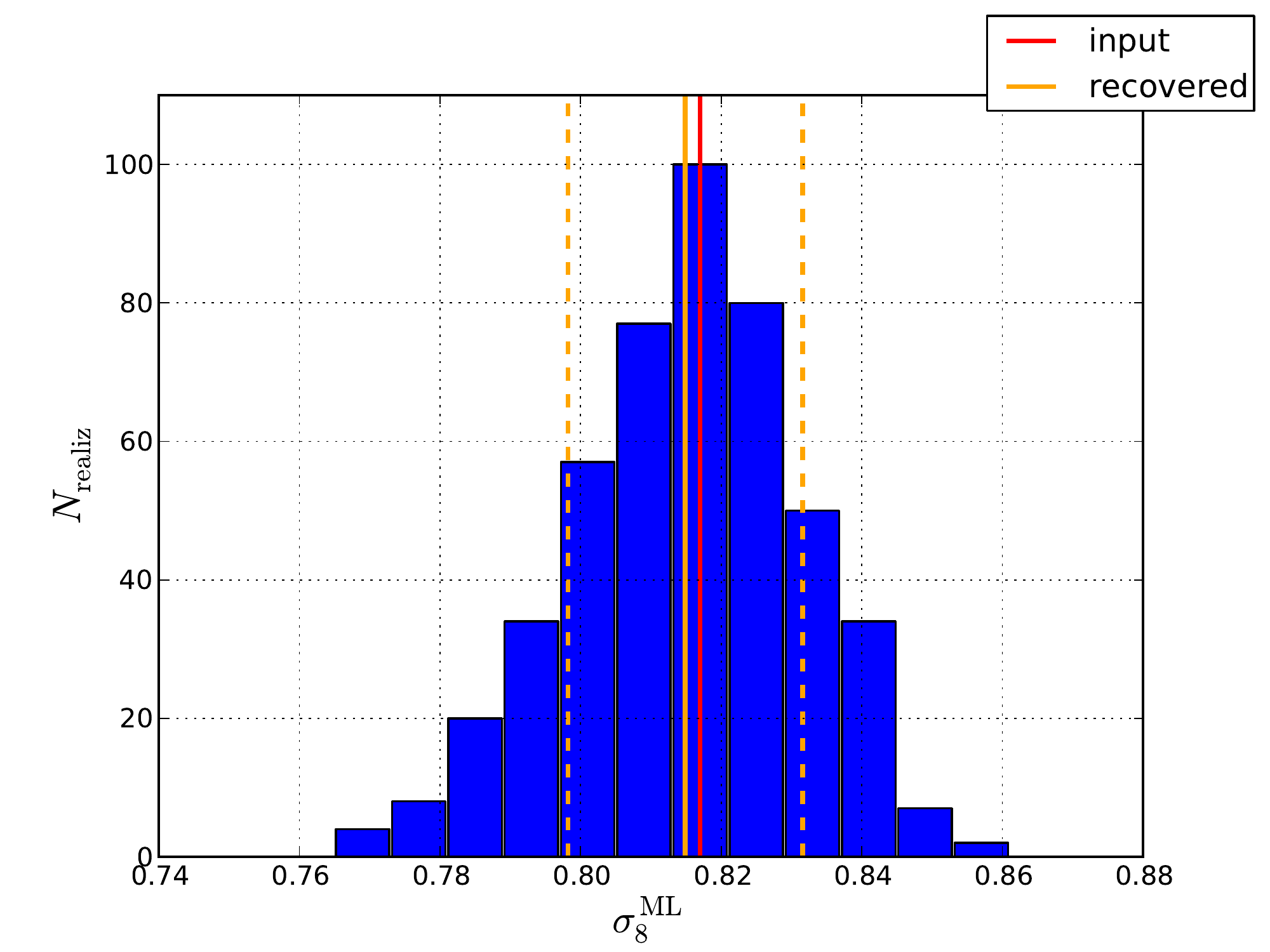}
\caption{Histogram of ML $\sigma_8$ values recovered from 476 Monte Carlo simulations processed through the likelihood in Eq.~(\ref{eq.like}) with $P_0$ fixed to unity, $\Omega_m$ fixed to 0.282, and the nuisance parameter $\sigma^2_{\rm nuis}$ marginalized over.  The input value, $\sigma_8 = 0.817$ (indicated by the red vertical line), is recovered in an unbiased fashion: the mean recovered value is $\langle \sigma_8^{\rm ML} \rangle = 0.815 \pm 0.017$ (indicated by the orange vertical lines).
\label{fig.sig8onlyhist}}
\end{figure}

We first set $P_0 = 1$ and $\Omega_m = 0.282$, and let only $\sigma_8$ vary.  To demonstrate the unbiased nature of the likelihood function, we use the Monte Carlo simulations described in Section~\ref{sec:sims}.  We place no prior on $\sigma_8$, and treat $\sigma^2_{\rm nuis}$ as described above, marginalizing over it in our results.  We process 476 Monte Carlo simulations for our fiducial model ($\sigma_8 = 0.817$, $\sigma^2_{\rm nuis} = \sigma^2_{\rm fid}$, $P_0 = 1$, $\Omega_m = 0.282$) through the likelihood.  We marginalize the likelihood over $\sigma^2_{\rm nuis}$ for each simulation and then find the maximum-likelihood (ML) value for $\sigma_8$.  The resulting histogram of ML $\sigma_8$ values is shown in Fig.~\ref{fig.sig8onlyhist}.  The mean recovered value is $\langle \sigma_8^{\rm ML} \rangle = 0.815 \pm 0.017$, which agrees with the input value $\sigma_8 = 0.817$.  This result indicates that the likelihood for $\sigma_8$ is unbiased.

\begin{figure}
\centering
\includegraphics[width=0.75\textheight]{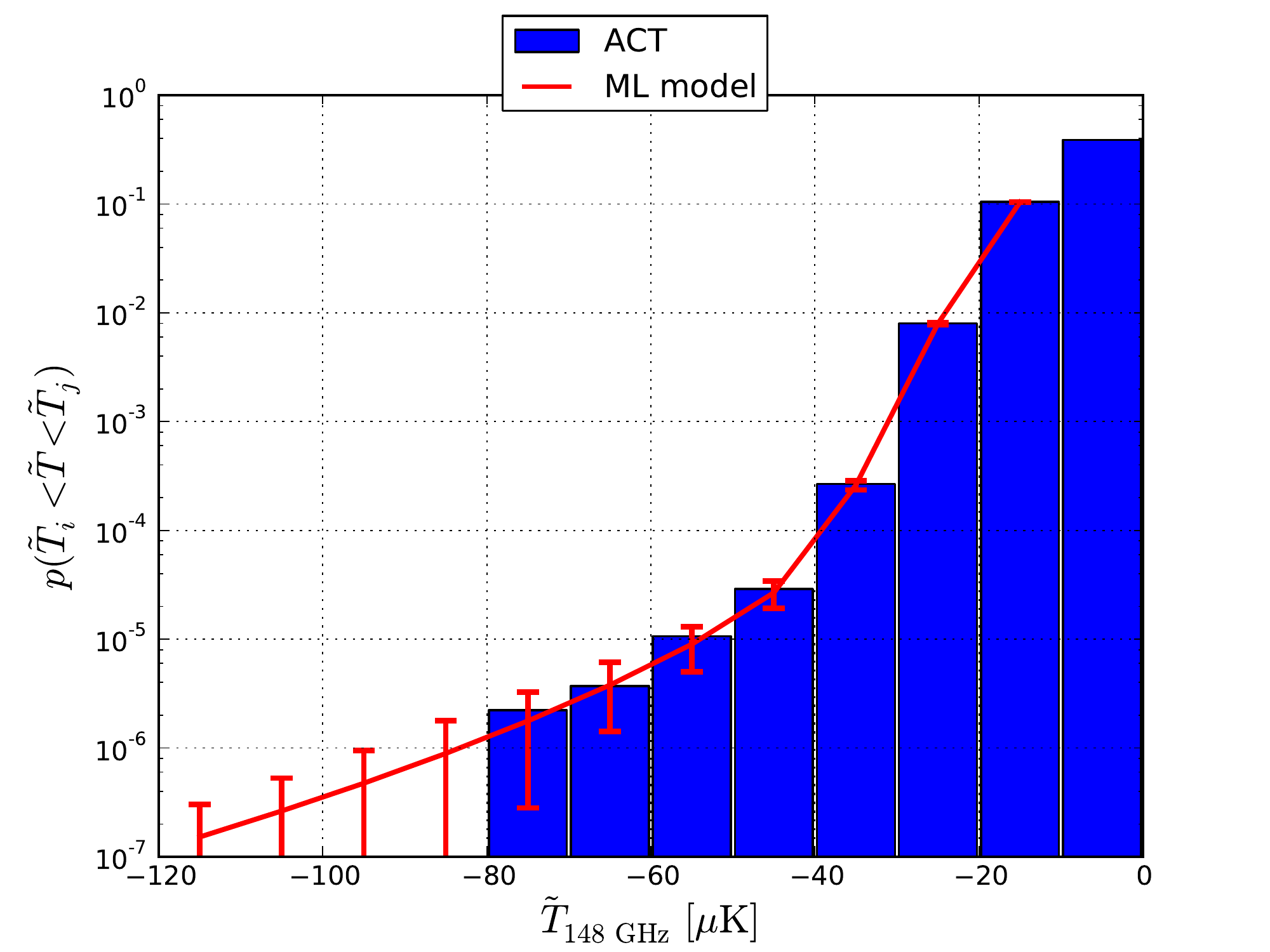}
\caption{Comparison of the measured ACT PDF to the ML PDF model.  The smallest-$|\tilde{T}|$ bin in the ML PDF model is not shown in this plot because it is not fit in the likelihood analysis.  The ML model provides a reasonable fit to the data, with $\chi^2 = 2.0$ for $5$ effective d.o.f. (see discussion of re-binning in the likelihood analysis in Section~\ref{sec:like}), corresponding to a PTE~$=0.85$.  The bin-to-bin correlations are strong, so the goodness-of-fit cannot be accurately assessed by eye.
\label{fig.PDFACTandML}}
\end{figure}

We then apply this likelihood function to the ACT data.  We find
\beq
\sigma_8 = 0.783 \pm 0.018 \,\,\,\,\,\,\, (P_0 = 1, \Omega_m = 0.282).
\label{eq.ACTsig8only}
\eeq
This result can be directly compared to the primary constraint derived from the tSZ skewness measurement using essentially the same data set in~W12, which found $\sigma_8 = 0.79 \pm 0.03$ under the same set of assumptions, i.e., assuming a fixed gas pressure profile model and holding all other cosmological parameters fixed (relativistic corrections and IR ``fill-in'' were neglected in~W12, but would lead to a shift in $\sigma_8$ significantly smaller than the statistical error bar).  The error bar on the PDF-derived result is roughly a factor of two smaller than that from the skewness alone.  The additional statistical power contained in the higher moments of the tSZ field beyond the skewness is responsible for this decrease in the error.  The central value of $\sigma_8$ is also slightly lower than that found in~W12.  The most plausible explanation for this result is the empty bins in the tail of the ACT PDF (see Figs.~\ref{fig.ACTPDF} and~\ref{fig.PDFACTandML}) --- the higher moments beyond the skewness are progressively more dominated by the bins far in the tail, and hence their influence likely pulls $\sigma_8$ down somewhat from the skewness-derived result.

At this level of precision on $\sigma_8$, it is important to consider uncertainties in the gas pressure profile model as well.  
We thus consider the case in which both $P_0$ and $\sigma_8$ are allowed to vary.  We place the same prior on $\sigma^2_{\rm nuis}$ as above and marginalize over it to obtain a 2D likelihood for $P_0$ and $\sigma_8$ ($\Omega_m = 0.282$ is still held fixed).  The parameter dependences shown in Fig.~\ref{fig.paramdep} indicate that the effects of $\sigma_8$ and $P_0$ on the tSZ PDF are not completely degenerate.  However, a full breaking of the degeneracy requires a measurement of the PDF over a wide range of $\tilde{T}$ values, so that the different changes in the shape of the PDF produced by the two parameters can be distinguished.  Testing the 2D likelihood for $P_0$ and $\sigma_8$ using our Monte Carlo simulations, we find that the degeneracy cannot be meaningfully broken with the ACT Equatorial data alone, especially due to the necessity of combining most of the bins in the tail of the PDF in order to ``Gaussianize'' the likelihood, as discussed in the previous section.  Thus, to obtain a meaningful marginalized 1D constraint on $\sigma_8$, we place a prior on $P_0$ motivated by the many recent observational studies that agree with our fiducial pressure profile (the ``AGN feedback'' fit from~\cite{Battagliaetal2012})~\cite{Arnaudetal2010,Sunetal2011,Planck2013stack,Planck2013ymap,Hajianetal2013,Hill-Spergel2014}.  We place a Gaussian prior of width $0.1$ on $P_0$, centered on the fiducial value $P_0 = 1$.  This procedure is analogous to that used in~\cite{Planck2013counts}, in which a flat prior between $[0.7,1.0]$ is placed on the ``hydrostatic mass bias'' $(1-b)$.  Our central value of $P_0 = 1$ corresponds roughly to $(1-b) \approx 0.85$--$0.95$, as discussed in Section~\ref{sec:theorymodel}, making our analysis quite similar to~\cite{Planck2013counts}.

Marginalizing the 2D likelihood over $P_0$ with this prior, we find
\beq
\sigma_8 = 0.781 \pm 0.025 \,\,\,\,\,\,\, (P_0 \,\, \mathrm{marg.}, \Omega_m = 0.282).
\label{eq.ACTsig8P0marg}
\eeq
Comparing these error bars to those in Eq.~(\ref{eq.ACTsig8only}), the inferred contribution to the overall error arising from the ICM physics uncertainty is $\sigma^{\rm ICM} ( \sigma_8 ) = 0.017$.  This value is comparable to the statistical error in Eq.~(\ref{eq.ACTsig8only}), demonstrating that further independent constraints on the ICM pressure profile would be beneficial in tightening the error on $\sigma_8$.

Finally, we allow $\Omega_m$ to vary in the likelihood as well.  The PDF data alone cannot constrain $\sigma_8$ and $\Omega_m$ simultaneously, so we focus on the best-determined degenerate combination of these parameters, $\Sigma_8 \equiv \sigma_8 \left( \Omega_m/0.282 \right)^{0.2}$.  Analogous to the $\sigma_8$ results above, we consider cases where $P_0$ is either held fixed to unity or allowed to vary and subsequently marginalized over.  In the former case, we find
\beq
\Sigma_8 = 0.783 \pm 0.019 \,\,\,\,\,\,\, (P_0 = 1),
\label{eq.ACTS8only}
\eeq
while in the latter case, we find
\beq
\Sigma_8 = 0.779^{+0.026}_{-0.025} \,\,\,\,\,\,\, (P_0 \,\, \mathrm{marg.}).
\label{eq.ACTS8P0marg}
\eeq
Comparing these error bars to those in Eq.~(\ref{eq.ACTS8only}), the inferred contribution to the overall error arising from the ICM physics uncertainty is $\sigma^{\rm ICM} ( \Sigma_8 ) = {}^{+0.018}_{-0.016}$, essentially identical to that found for $\sigma_8$ above.

\begin{figure}
\centering
\includegraphics[totalheight=0.5\textheight]{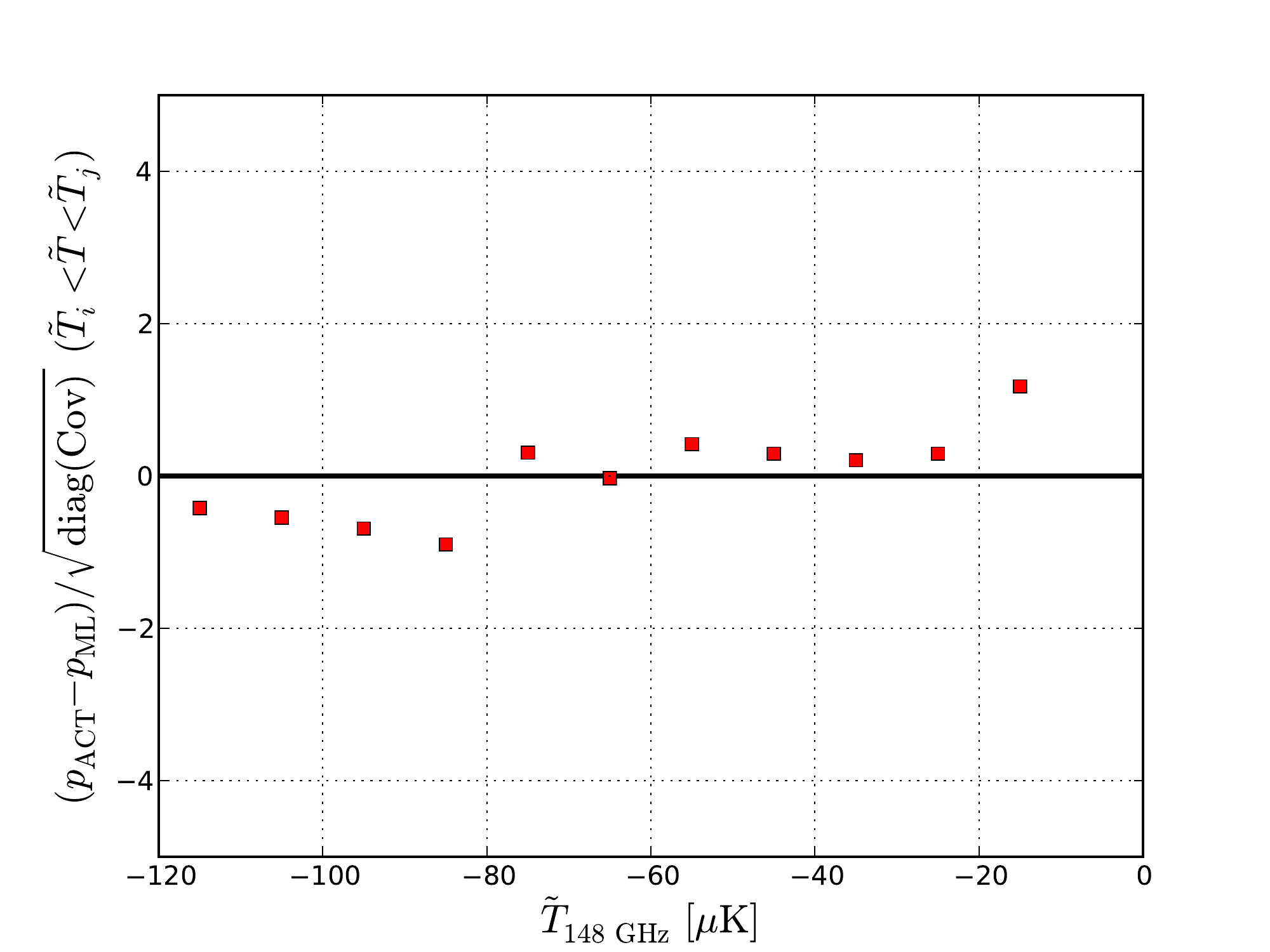}
\caption{Difference between the measured ACT PDF and the ML model divided by the square root of the diagonal elements of the covariance matrix of the ML model.  This plot obscures the significant bin-to-bin correlations in the PDF (see Fig.~\ref{fig.PDFcovfid}), and thus caution must be used in any visual interpretation.  However, the plot suggests that the empty outermost bins in the tail of the ACT PDF (i.e., the lack of extremely massive clusters in the observed field compared to the Tinker mass function prediction) are responsible for pulling the best-fit amplitude of the ML model down from the value that would likely be preferred by the populated bins.
\label{fig.PDFACTandMLratiosig}}
\end{figure}

Fig.~\ref{fig.PDFACTandML} presents a comparison of the ACT PDF to the ML PDF model, which is located at $\{ \sigma_8,\Omega_m,P_0 \} = \{ 0.765, 0.292, 1.05 \}$ (i.e., this is the global ML point in the full parameter space, but due to degeneracies between these parameters the precise values here are not particularly informative).  We leave out the smallest-$|\tilde{T}|$ bin in the ML PDF model curve in this plot because it is not fit in the likelihood analysis.  
The comparison is investigated further in Fig.~\ref{fig.PDFACTandMLratiosig}, which shows the difference between the ACT PDF and the ML model divided by the square root of the diagonal elements of the covariance matrix of the ML model.  We emphasize that the diagonal-only nature of this plot obscures the significant bin-to-bin correlations in the PDF, and thus no strong inferences should be visually derived.  Nonetheless, it is interesting that the best-fit model lies slightly high over most of the range of the PDF, with the large-$|\tilde{T}|$ bins apparently responsible for the pull toward lower amplitudes, especially because of their strong dependence on $\sigma_8$ (see Fig.~\ref{fig.paramdep}).  Finally, we assess the goodness-of-fit of the ML PDF model by computing $\chi^2 = ( p_i^{\rm ML} - \hat{p}_i) (\mathrm{Cov}_{\rm ML}^{-1})_{ij} ( p_j^{\rm ML} - \hat{p}_j)$, where we also use the covariance matrix corresponding to the ML model.  We find $\chi^2 = 2.0$ for $5$ effective d.o.f.~(the number of d.o.f.~left after the re-binning described in the previous section), which corresponds to a probability-to-exceed (PTE) of $0.85$.  Thus, the ML PDF model provides a reasonable fit to the observed ACT PDF.

\subsection{IR ``Fill-In''}
\label{sec:CIBbias}

A currently open question in tSZ analyses is the extent to which IR emission from dusty, star-forming galaxies ``fills in'' the tSZ decrements measured at $\approx 150$ GHz by ACT, SPT, and other experiments.  This effect is distinct from the CIB signal in the 148 GHz PDF described in Section~\ref{sec:noisyPDF}, which arises from diffuse CIB emission; we negate the diffuse CIB signal by considering only the $\tilde{T} < 0$ region of the ACT PDF, and by further discarding the $[-10\, \mu {\rm K}, 0\, \mu {\rm K}]$ bin in the likelihood analysis (see Section~\ref{sec:like}).  The ``fill-in'' effect rather arises from the small fraction of IR sources that may lie in the clusters comprising the tSZ PDF signal.  Since this IR emission is positive at 148 GHz while the tSZ signal is negative, the IR sources can ``fill in'' the tSZ decrement, thus biasing any inferred constraints from the measurement.  This effect is analogous to that arising from radio sources, which has been shown to be negligible for the typical massive tSZ cluster at 150 GHz --- for example, Sayers et al.~\cite{Sayersetal2013} find $<1\%$ contamination in nearly all clusters in a sample of moderate-redshift, fairly massive systems studied with Bolocam (see also~\cite{Diego-Partridge2010}).  Low frequency-based studies have found that radio sources are even less prevalent in lower mass systems (e.g.,~\cite{Bestetal2007,Grallaetal2011}).Ê Observations of radio sources in galaxy clusters find no correlation between their spectral properties and the cluster mass~\cite{Linetal2009}, so the low frequency-based studies imply that radio sources are not likely to be a significant contaminant in low-mass clusters.Ê Indeed,~\cite{Linetal2009} calculate the fraction of contaminated $M \approx 10^{14} \, M_{\odot}/h$ clusters to be $<10$\% at $z = 0.1$, and $\lsim 2$\% at higher redshifts.  Thus, we will consider only contaminating emission from IR sources in the following analysis.

For a single-frequency measurement, IR (or radio) ``fill-in'' effects are degenerate with a change in the behavior of the gas pressure profile (e.g., the IR emission would bias $P_0 < 1$ in our model).  Also, although this effect is related to the tSZ -- CIB correlation currently being studied in CMB power spectrum measurements~\cite{Addisonetal2012,Sieversetal2013,Georgeetal2014}, it is not precisely identical, since in our case only the one-halo term of the correlation is relevant.  Multi-frequency measurements have the possibility to disentangle the IR emission from the tSZ signal, and recently this issue has come under consideration using ACT~\cite{Lindneretal2014} and SPT data~\cite{Liuetal2014}.  To address this issue in our tSZ PDF analysis, we consider the results from these studies and make independent measurements using the ACT 218 GHz maps.  In particular, we stack the ACT 218 GHz maps at the locations of the pixels that comprise the $\tilde{T}_{148\, {\rm GHz}} < -41.5\, \mu {\rm K}$ region of the 148 GHz PDF, corresponding to decrements greater than $5\sigma$, where $\sigma$ is the map rms.  We process and filter the 218 GHz maps in the same way as the 148 GHz maps (see Section~\ref{sec:data}) and then compute the average temperature at these pixel locations.   

Such 218 GHz stacking measurements need to be corrected for a CMB-induced bias, as there are common CMB fluctuations at the two frequencies despite the filtering. We perform this calculation using the 148 GHz simulations described in Section~\ref{sec:sims} as well as 90 additional noisy 218 GHz simulations with the same underlying CMB realizations, finding a bias of $-2.6\, \mu {\rm K}$ at 218 GHz. 
Furthermore, just like this CMB bias, a bias arises from the uncorrelated CIB component, which is also common across frequencies and exists both at 148 and 218 GHz. Using approximate Gaussian simulations, we correct for it as well.  We then re-scale the final, corrected result from 218 GHz to 148 GHz (roughly a factor of $0.3$) and obtain a IR ``fill-in'' bias of $\tilde{T}^{\rm IR}_{148\, {\rm GHz}} = 0.1 \pm 1.2\, \mu {\rm K}$ for $\tilde{T}_{148\, {\rm GHz}} < -41.5\, \mu {\rm K}$.  This result indicates that for the massive clusters comprising the tail of the ACT tSZ PDF ($M \gtrsim 6 \times 10^{14} \, M_{\odot}/h$ --- see Fig.~\ref{fig.masscontrib}), the IR ``fill-in'' bias is consistent with zero in our data.  This result is also consistent with the results from~\cite{Lindneretal2014} focusing on emission from sub-millimeter galaxies ``filling in'' the tSZ decrements in the massive clusters in the ACT southern strip.

However, the clusters sourcing the moderate decrements in the ACT tSZ PDF ($-41.5\, \mu {\rm K} < \tilde{T}_{148\, {\rm GHz}} < -20\, \mu {\rm K}$) could still be affected by IR emission.  A stacking analysis on the pixels comprising these decrements is complicated by the fact that many are simply noise, and in fact our formalism cannot unambiguously distinguish the origin of pixels in this region as ``true'' tSZ or noise (see Eq.~(\ref{eq.yPDF}) and Fig.~\ref{fig.PDFfid}).  In contrast, the stacking of the 218 GHz maps on the tSZ-dominated pixels in the PDF tail is robust and the physical interpretation is unambiguous.  To estimate the residual bias from IR emission in the moderate-decrement region, we build a simple model from the results found above and in the recent SPT stacking analysis on a sample of X-ray-selected objects~\cite{Liuetal2014}.  The SPT analysis finds an IR ``fill-in'' of $(32 \pm 18)$\% at 150 GHz for a sample with median mass $M_{500} \approx 1.5 \times 10^{14} \, M_{\odot}$ (corresponding roughly to a virial mass $M \approx 2 \times 10^{14} \, M_{\odot}/h$ over the redshift range of interest).  However, additional SPT and Herschel results from~\cite{Bleem2013} indicate that the ``fill-in'' bias declines rather rapidly with increasing cluster mass to $(5 \pm 5)$\% at mass scales $M \approx (2.5$--$5) \times 10^{14} \, M_{\odot}/h$.  Moreover, our results above as well as those in~\cite{Lindneretal2014} indicate a ``fill-in'' bias of at most $\approx 2$--$3$\% at higher masses.

\begin{figure}
\centering
\includegraphics[totalheight=0.5\textheight]{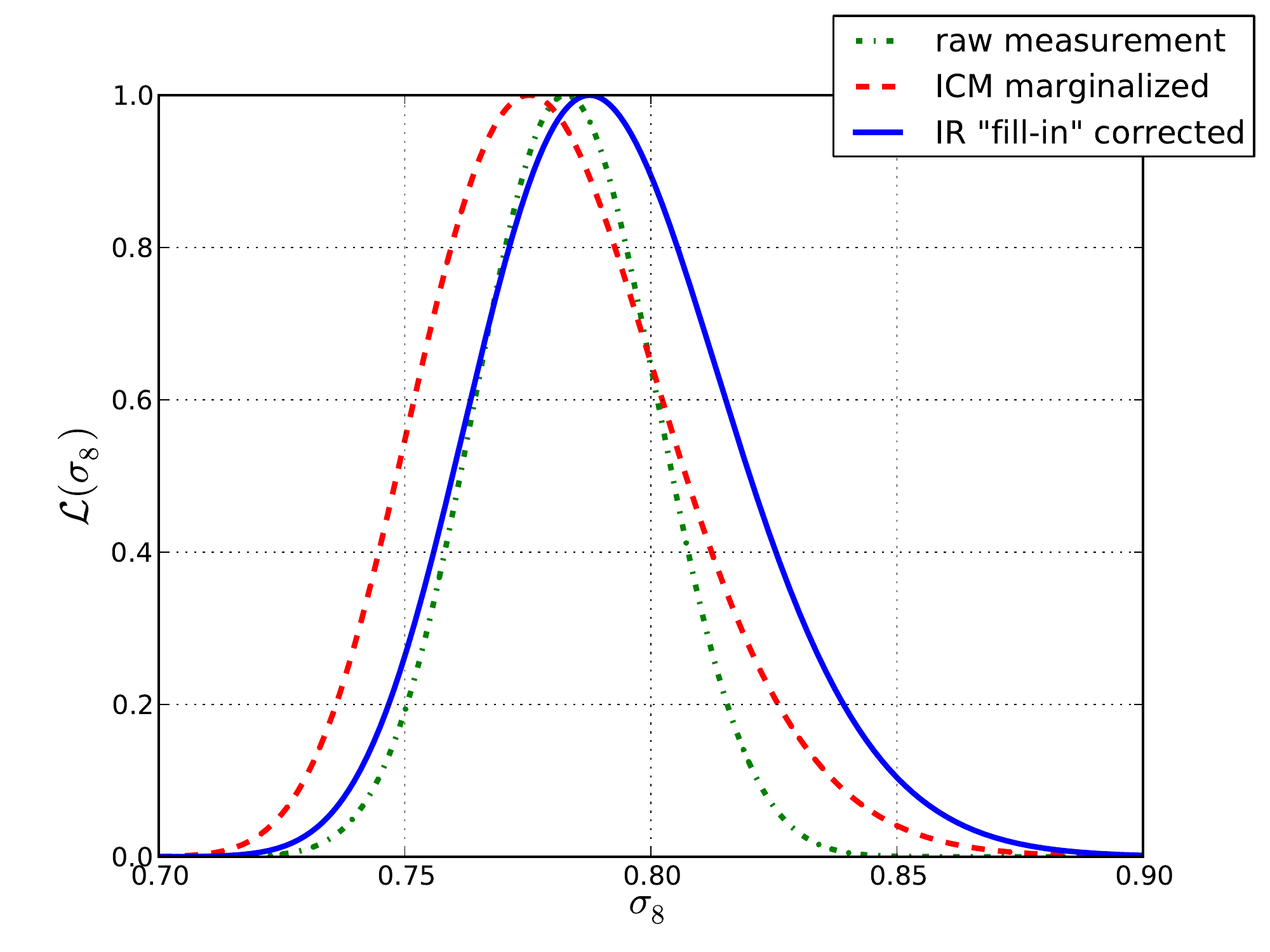}
\caption{Likelihood for $\sigma_8$ computed from the ACT 148 GHz PDF using the likelihood function described in Section~\ref{sec:like}.  The dash-dotted green curve shows the raw constraint with the normalization of the pressure--mass relation $P_0$ fixed to unity and $\Omega_m$ fixed to $0.282$ (i.e., only $\sigma_8$ and the non-tSZ foreground amplitude $\sigma^2_{\rm nuis}$ are free to vary, and the latter is marginalized over).  The dashed red curve shows the constraint when $P_0$ is also allowed to vary and is subsequently marginalized over, thus propagating the ICM physics uncertainty into the $\sigma_8$ likelihood.  The blue solid curve shows the final result after correcting for the IR ``fill-in'' bias described in Section~\ref{sec:CIBbias}.  The constraints on $\Sigma_8$ are nearly identical to those shown for $\sigma_8$ here, so we do not plot them for clarity.
\label{fig.ACTsig8}}
\end{figure}

We thus consider a simple mass-dependent model for the IR emission that satisfies these constraints in the high- and low-mass regimes. We set the ``fill-in'' value to $15$\% at the lowest-mass end in our theoretical calculations ($M = 2 \times 10^{14} \, M_{\odot}/h$) and then assume the IR emission scales as $M$, which means the ``fill-in'' effect scales roughly as $M^{-2/3}$ since the tSZ signal scales roughly as $M^{5/3}$. This model is overly simplistic, but is not dissimilar from the assumptions made in the simulations of~\cite{Sehgaletal2010}, based on the CIB model of~\cite{Righietal2008}.  The simulations assume the number of IR sources $N_{\rm gal} \propto M$ above a threshold mass (which we are well above in our analysis), and then model the IR emission as $N_{\rm gal}$ times luminosity, with a realistic luminosity prescription, whereas our crude approach essentially assumes all IR sources have identical luminosity.  We implement this model in the simulations described in Section~\ref{sec:sims} and then process the resulting PDFs through the likelihood described in Section~\ref{sec:like}.  The IR ``fill-in'' induces a bias $\Delta \sigma_8^{\rm IR} = -0.012$ on $\sigma_8$ and a bias $\Delta \Sigma_8^{\rm IR} = -0.011$ on $\Sigma_8$, which are approximately equal to the biases one finds when assuming a constant $10$\% IR ``fill-in'' of the tSZ decrement for clusters of all masses.

Thus, we must correct the constraints from the previous section for these biases.  We use the above analysis for the fiducial correction and assign an uncertainty of $50$\% on the correction itself.  Hence, Eq.~(\ref{eq.ACTsig8P0marg}) becomes
\beq
\sigma_8 = 0.793 \pm 0.018\, ({\rm stat.})\, \pm 0.017\, ({\rm ICM\,\, syst.})\, \pm 0.006\, ({\rm IR\,\, syst.}) \,,
\label{eq.ACTsig8P0margCIBcorr}
\eeq
and Eq.~(\ref{eq.ACTS8P0marg}) becomes
\beq
\Sigma_8 = 0.790 \pm 0.019\, ({\rm stat.})\, {}^{+0.018}_{-0.016} \, ({\rm ICM\,\, syst.})\, \pm 0.006\, ({\rm IR\,\, syst.}) \,,
\label{eq.ACTS8P0margCIBcorr}
\eeq
where we have explicitly separated the contributions to the error from statistical uncertainty, ICM systematic uncertainty as derived in Eqs.~(\ref{eq.ACTsig8P0marg}) and~(\ref{eq.ACTS8P0marg}), and IR ``fill-in'' systematic uncertainty, respectively.  Finally, we note that in principle the IR ``fill-in'' bias should have been corrected for in the tSZ skewness analysis of~W12, but it is subdominant to the statistical error bar on $\sigma_8$ in that study.  The improved precision of our constraint from the tSZ PDF necessitates the more careful consideration presented here.

\subsection{Discussion}

The 1D likelihoods for $\sigma_8$ corresponding to the constraints in Eqs.~(\ref{eq.ACTsig8only}),~(\ref{eq.ACTsig8P0marg}), and~(\ref{eq.ACTsig8P0margCIBcorr}) are shown in Fig.~\ref{fig.ACTsig8}.  The likelihoods for $\Sigma_8$ in Eqs.~(\ref{eq.ACTS8only}),~(\ref{eq.ACTS8P0marg}), and~(\ref{eq.ACTS8P0margCIBcorr}) are nearly identical to those shown in Fig.~\ref{fig.ACTsig8}.  Our final result for $\sigma_8$ is consistent with W12.  In addition, although we cannot make a completely like-for-like comparison to the ACT cluster count analysis of~\cite{Hasselfieldetal2013} due to the different ICM physics models considered, we can consider the result for $\sigma_8$ found in that analysis when using the ``universal pressure profile'' model of~\cite{Arnaudetal2010}, which is the most similar model to that used here~\cite{Battagliaetal2012}.  Given this ICM model, the cluster count analysis found $\sigma_8 = 0.786 \pm 0.013$ (for ACT clusters + WMAP7 CMB data), which agrees with our result.  This error bar cannot be compared directly to the errors quoted in our PDF analysis, since it also includes WMAP7 CMB data, which provide additional constraining power.  A more useful error bar comparison is to the fixed-scaling-relation ``BBN+H0+ACTcl(B12)'' results (see Table 3 of~\cite{Hasselfieldetal2013}), which do not use CMB data.  The resulting constraint, $\sigma_8 (\Omega_m / 0.282)^{0.3} = 0.837 \pm 0.032$, can be compared to the statistical error of $\pm 0.019$ that we find on $\Sigma_8$ in Eq.~(\ref{eq.ACTS8only}) --- note that the best-constrained degenerate combination is slightly different for the counts than for the PDF.  This comparison demonstrates the additional constraining power arising from the low-SNR objects in the PDF.  When considering the results in~\cite{Hasselfieldetal2013} based on dynamical mass measurements, the $\sigma_8$ constraint ($\sigma_8 = 0.829 \pm 0.024$ for ACT clusters + WMAP7 CMB data) is consistent with our constraint from the tSZ PDF as well.  The agreement between the ACT PDF and cluster count analyses is noteworthy given that the methods are sensitive to somewhat different populations of clusters and involve different treatments of systematic effects.  

Our results are also consistent with other recent constraints on $\sigma_8$ from tSZ and galaxy cluster measurements.  In addition to~W12, the most closely related analysis to that presented here is the SPT tSZ bispectrum measurement, which found $\sigma_8 = 0.787 \pm 0.031$~\cite{Crawfordetal2014}.  The error bar on $\sigma_8$ in~\cite{Crawfordetal2014} is dominated by their assumed $36$\% ICM modeling uncertainty on the tSZ bispectrum amplitude, which is nearly identical to our assumed $10$\% uncertainty on $P_0$, since the bispectrum amplitude goes as $P_0^3$.  In addition to the ACT cluster counts mentioned above~\cite{Hasselfieldetal2013}, our results are consistent with cluster count analyses using SPT data~\cite{Reichardtetal2013}, \emph{Planck} data~\cite{Planck2013counts}, and X-ray data~\cite{Vikhlininetal2009}.  The SPT cluster counts constrained $\sigma_8 = 0.798 \pm 0.017$ (including WMAP7 and SPT CMB data to break degeneracies), while the \emph{Planck} cluster counts constrained $\sigma_8 = 0.77 \pm 0.02$ (including BAO-derived priors to break degeneracies).  Note that the various analyses make somewhat different assumptions about the ICM physics.  Our results are also consistent with constraints on $\sigma_8$ from the high-$\ell$ tSZ power spectrum measured by ACT~\cite{Sieversetal2013} and SPT~\cite{Georgeetal2014}, although these constraints are more sensitive to ICM physics uncertainty than other tSZ probes because of the lower-mass, higher-redshift cluster population sourcing the signal.  We also note consistency with results from the tSZ power spectrum measured by \emph{Planck}~\cite{Planck2013ymap}, which found $\sigma_8 (\Omega_m/0.28)^{0.4} = 0.784 \pm 0.016$ (assuming no ICM physics uncertainty).  Finally, our results are consistent with constraints from two tSZ cross-correlation analyses: the tSZ -- CMB lensing cross-power spectrum measured using \emph{Planck} data in~\cite{Hill-Spergel2014}, which found $\sigma_8 (\Omega_m/0.282)^{0.26} = 0.824 \pm 0.029$ (under the assumption of a fixed ICM model), and the tSZ -- X-ray cross-power spectrum measured using \emph{Planck}, WMAP, and ROSAT data in~\cite{Hajianetal2013}, which found $\sigma_8 (\Omega_m/0.282)^{0.26} = 0.81 \pm 0.02$.

A common thread among many of these measurements is a preference for a somewhat lower value of $\sigma_8$ than that preferred by the primordial CMB, as has been noted elsewhere.  For example, WMAP9 found $\sigma_8 = 0.821 \pm 0.023$~\cite{Hinshawetal2013}, \emph{Planck} + WMAP polarization found $\sigma_8 = 0.829 \pm 0.012$~\cite{Planck2013params}, and a separate analysis of \emph{Planck} +WMAP polarization found $\sigma_8 = 0.817 \pm 0.012$~\cite{Spergeletal2013}.  Resolving this discrepancy between low- and high-redshift measurements of the amplitude of density fluctuations is an important goal, with new physics (e.g., massive neutrinos~\cite{Planck2013counts}), a better understanding of the ICM, or currently neglected systematics providing possible solutions.  A key parameter in the CMB analyses is $\tau$, the optical depth to reionization, which is degenerate with the scalar amplitude $A_s$ ($A_s$ can be mapped onto $\sigma_8$ if all other parameters are held fixed).  Current analyses rely almost entirely on WMAP EE data to determine $\tau$, but a polarized dust cleaning analysis using \emph{Planck} 353 GHz data suggests the WMAP measurement may be biased $\approx 1\sigma$ high~\cite{Planck2013likelihood} (see their Appendix E).  If $\tau$ shifts lower by $1\sigma$, then the CMB-inferred $\sigma_8$ would also shift lower by $\approx 1\sigma$, reducing the discrepancy with low-redshift measurements.  Regardless of the CMB developments, further calibration of the ICM physics using tSZ, X-ray, and gravitational lensing data will be essential to better characterize the low-redshift amplitude of fluctuations.

\section{Conclusion and Outlook}
\label{sec:outlook}

In this paper we have presented the first application of the one-point PDF to statistical measurements of the tSZ effect.  After introducing a theoretical framework in which to compute the noise-convolved, observable tSZ PDF and its covariance matrix, we measure the signal in the ACT 148 GHz Equatorial data and use it to constrain $\sigma_8$.  Directly comparing the fixed-ICM-physics result to our earlier tSZ skewness analysis in~W12 indicates that the PDF decreases the error bar on $\sigma_8$ by a factor of $\approx 2$, using effectively the same data set.  This result clearly demonstrates the additional statistical power in the higher moments present in the PDF.  The primary theoretical uncertainty in the PDF modeling arises from the ICM gas pressure profile --- the halo mass function is well-determined over the mass range that dominates our constraints~\cite{Tinkeretal2008}.  In principle, the tSZ PDF allows for the breaking of the ICM--cosmology degeneracy, although the current data and likelihood analysis are not quite sufficient to realize this goal.  However, the PDF framework allows for simple propagation of the uncertainty on parameters describing the ICM pressure--mass relation into our constraint on $\sigma_8$.  

A number of clear extensions to this approach can be implemented in the near future.  Of particular interest is the 2D joint tSZ--weak lensing PDF, $p(T,\kappa)$, which should strongly break the degeneracy between cosmological and ICM parameters present in tSZ statistics.  Effectively, the weak lensing information provides information on the mass scale contributing to a given tSZ signal.  This allows simultaneous determination of the pressure--mass relation and cosmological parameters at a level closer to the full forecasted precision of the tSZ statistics.  Another extension of interest is the measurement of the tSZ PDF after applying different $\ell$-space filters, as opposed to the single-filter approach used in this work.  By measuring the PDF as a function of filter scale, it should be possible to recover much of the angular information lost by using only the zero-lag moments of the tSZ field rather than the full polyspectra.  This angular information then provides constraints on the structure of the ICM pressure profile.  We leave these considerations for future work, in anticipation of upcoming, higher-SNR measurements of the tSZ PDF.

\begin{acknowledgments}
ACT operates in the Parque Astron\'{o}mico Atacama in northern Chile under the auspices of Programa de Astronom\'{i}a, a program of the Comisi\'{o}n Nacional de Investigaci\'{o}n Cient\'{i}fica y Tecnol\'{o}gica de Chile (CONICYT).  This work was supported by the U.S.\ NSF through awards AST-0408698, PHY-0355328, AST-0707731 and PIRE-0507768, as well as by Princeton Univ., the Univ.~of Pennsylvania, FONDAP, Basal, Centre AIUC, RCUK Fellowship (JD), CONICYT grants QUIMAL-120001 and FONDECYT-1141113 (RD), NASA grant NNX08AH30G (AH), NSERC PGSD (ADH), ERC grant 259505 (EC, JD), a NASA Space Technology Research Fellowship (BS), NSF AST-1108790 (AK), NASA grant NNX14AB57G, and NASA Theory Grant NNX12AG72G and NSF AST-1311756 (JCH, DNS).  JCH gratefully acknowledges support from the Simons Foundation through the Simons Society of Fellows.  Research at Perimeter Institute is supported by the Government of Canada through Industry Canada and by the Province of Ontario through the Ministry of Research \& Innovation.  KMS was supported by an NSERC Discovery Grant.Ê  We thank  B.\ Berger, R.\ Escribano, T.\ Evans, D.\ Faber, P.\ Gallardo, A.\ Gomez, M.\ Gordon, D.\ Holtz, M.\ McLaren, W.\ Page, R.\ Plimpton, D.\ Sanchez, O.\ Stryzak, M.\ Uehara, and Astro-Norte for assistance with ACT.  Computations were performed on the GPC supercomputer at the SciNet HPC consortium.  SciNet is funded by the Canada Foundation for Innovation under the auspices of Compute Canada, the Government of Ontario, the Ontario Research Fund --- Research Excellence, and the Univ.~of Toronto.
\end{acknowledgments}


\begin{thebibliography}{3}

\bibitem[Swetz et al.(2011)]{Swe11} Swetz, D.~S., Ade, P.~A.~R., Amiri, M., et al.\ 2011, \apjs, 194, 41 
\bibitem[Niemack et al.(2010)]{Niemacketal2010} Niemack, M.~D., Ade, P.~A.~R., Aguirre, J., et al.\ 2010, \procspie, 7741
\bibitem[Carlstrom et al.(2011)]{Car11} Carlstrom, J.~E., Ade, P.~A.~R., Aird, K.~A., et al.\ 2011, PASP, 123, 568 
\bibitem[Austermann et al.(2012)]{Austermannetal2012} Austermann, J.~E., Aird, K.~A., Beall, J.~A., et al.\ 2012, \procspie, 8452
\bibitem[Planck Collaboration et al.(2013)]{Planck2013overview} Planck Collaboration, Ade, P.~A.~R., Aghanim, N., et al.\ 2013, arXiv:1303.5062 
\bibitem[Kermish et al.(2012)]{Kermishetal2012} Kermish, Z.~D., Ade, P., Anthony, A., et al.\ 2012, \procspie, 8452
\bibitem[Zeldovich \& Sunyaev(1969)]{Zel69} Zel'dovich, Y.~B., \& Sunyaev, R.~A.\ 1969, \apss, 4, 301
\bibitem[Sunyaev \& Zeldovich(1970)]{Sunyaev-Zeldovich1970} Sunyaev, R.~A., \& Zeldovich, Y.~B.\ 1970, \apss, 7, 3 
\bibitem[Hand et al.(2012)]{Handetal2012} Hand, N., Addison, G.~E., Aubourg, E., et al.\ 2012, \prl, 109, 041101 
\bibitem[Hasselfield et al.(2013)]{Hasselfieldetal2013} Hasselfield, M., Hilton, M., Marriage, T.~A., et al.\ 2013, \jcap, 7, 8
\bibitem[Gralla et al.(2014)]{Grallaetal2014} Gralla, M.~B., Crichton, D., Marriage, T.~A., et al.\ 2014, \mnras, 445, 460 
\bibitem[Reichardt et al.(2013)]{Reichardtetal2013} Reichardt, C.~L., Stalder, B., Bleem, L.~E., et al.\ 2013, \apj, 763, 127 
\bibitem[Planck Collaboration et al.(2013)]{Planck2013counts} Planck Collaboration, Ade, P.~A.~R., Aghanim, N., et al.\ 2013, arXiv:1303.5080
\bibitem[Planck Collaboration et al.(2013)]{Planck2013LBG} Planck Collaboration, Ade, P.~A.~R., Aghanim, N., et al.\ 2013, \aap, 557, A52 
\bibitem[Sievers et al.(2013)]{Sieversetal2013} Sievers, J.~L., Hlozek, R.~A., Nolta, M.~R., et al.\ 2013, \jcap, 10, 60 
\bibitem[Story et al.(2013)]{Storyetal2013} Story, K.~T., Reichardt, C.~L., Hou, Z., et al.\ 2013, \apj, 779, 86 
\bibitem[Wilson et al.(2012)]{Wilsonetal2012} Wilson, M.~J., Sherwin, B.~D., Hill, J.~C., et al.\ 2012, \prd, 86, 122005
\bibitem[Crawford et al.(2014)]{Crawfordetal2014} Crawford, T.~M., Schaffer, K.~K., Bhattacharya, S., et al.\ 2014, \apj, 784, 143 
\bibitem[Planck Collaboration et al.(2013)]{Planck2013ymap} Planck Collaboration, Ade, P.~A.~R., Aghanim, N., et al.\ 2013, arXiv:1303.5081
\bibitem[Komatsu \& Seljak(2002)]{Komatsu-Seljak2002} Komatsu, E., \& Seljak, U.\ 2002, \mnras, 336, 1256 
\bibitem[Hill \& Sherwin(2013)]{Hill-Sherwin2013} Hill, J.~C., \& Sherwin, B.~D.\ 2013, \prd, 87, 023527
\bibitem[Hill \& Pajer(2013)]{Hill-Pajer2013} Hill, J.~C., \& Pajer, E.\ 2013, \prd, 88, 063526 
\bibitem{Bhattacharyaetal2012} Bhattacharya, S., Nagai, D., Shaw, L., Crawford, T., \& Holder, G.~P.\ 2012, \apj, 760, 5 
\bibitem[Shaw et al.(2010)]{Shawetal2010} Shaw, L.~D., Nagai, D., Bhattacharya, S., \& Lau, E.~T.\ 2010, \apj, 725, 1452
\bibitem[Battaglia et al.(2012)]{Battagliaetal2012} Battaglia, N., Bond, J.~R., Pfrommer, C., \& Sievers, J.~L.\ 2012, \apj, 758, 75 
\bibitem[Komatsu \& Kitayama(1999)]{Komatsu-Kitayama1999} Komatsu, E., \& Kitayama, T.\ 1999, \apjl, 526, L1
\bibitem[Hill \& Spergel(2014)]{Hill-Spergel2014} Hill, J.~C., \& Spergel, D.~N.\ 2014, \jcap, 2, 30
\bibitem[Van Waerbeke et al.(2014)]{vanWaerbekeetal2014} Van Waerbeke, L., Hinshaw, G., \& Murray, N.\ 2014, \prd, 89, 023508 
\bibitem[Das et al.(2014)]{Dasetal2014} Das, S., Louis, T., Nolta, M.~R., et al.\ 2014, \jcap, 4, 14 
\bibitem[Trac et al.(2011)]{Tracetal2011} Trac, H., Bode, P., \& Ostriker, J.~P.\ 2011, \apj, 727, 94 
\bibitem[Addison et al.(2012)]{Addisonetal2012} Addison, G.~E., Dunkley, J., \& Spergel, D.~N.\ 2012, \mnras, 427, 1741 
\bibitem[Mesinger et al.(2012)]{Mesingeretal2012} Mesinger, A., McQuinn, M., \& Spergel, D.~N.\ 2012, \mnras, 422, 1403 
\bibitem[Jain \& Van Waerbeke(2000)]{Jain-vanWaerbeke2000} Jain, B., \& Van Waerbeke, L.\ 2000, \apjl, 530, L1 
\bibitem[Neyrinck(2014)]{Neyrinck2014} Neyrinck, M.~C.\ 2014, arXiv:1407.4815 
\bibitem[Kratochvil et al.(2010)]{Kratochviletal2010} Kratochvil, J.~M., Haiman, Z., \& May, M.\ 2010, \prd, 81, 043519 
\bibitem[Scheuer(1957)]{Scheuer1957} Scheuer, P.~A.~G.\ 1957, Proceedings of the Cambridge Philosophical Society, 53, 764 
\bibitem[Condon(1974)]{Condon1974} Condon, J.~J.\ 1974, \apj, 188, 279 
\bibitem[Bond \& Myers(1996)]{Bond-Myers1996} Bond, J.~R., \& Myers, S.~T.\ 1996, \apjs, 103, 63 
\bibitem[Springel et al.(2001)]{Springeletal2001} Springel, V., White, M., \& Hernquist, L.\ 2001, \apj, 549, 681 
\bibitem[Springel et al.(2001)]{Springeletal2001ERR} Springel, V., White, M., \& Hernquist, L.\ 2001, \apj, 562, 1086 
\bibitem[Refregier \& Teyssier(2002)]{Refregier-Teyssier2002} Refregier, A., \& Teyssier, R.\ 2002, \prd, 66, 043002 
\bibitem[Seljak et al.(2001)]{Seljaketal2001} Seljak, U., Burwell, J., \& Pen, U.-L.\ 2001, \prd, 63, 063001 
\bibitem[Zhang et al.(2002)]{Zhangetal2002} Zhang, P., Pen, U.-L., \& Wang, B.\ 2002, \apj, 577, 555 
\bibitem[da Silva et al.(2000)]{daSilvaetal2000} da Silva, A.~C., Barbosa, D., Liddle, A.~R., \& Thomas, P.~A.\ 2000, \mnras, 317, 37 
\bibitem[Cooray(2000)]{Cooray2000} Cooray, A.\ 2000, \prd, 62, 103506 
\bibitem[Hinshaw et al.(2013)]{Hinshawetal2013} Hinshaw, G., Larson, D., Komatsu, E., et al.\ 2013, \apjs, 208, 19 
\bibitem[Planck Collaboration et al.(2013)]{Planck2013params} Planck Collaboration, Ade, P.~A.~R., Aghanim, N., et al.\ 2013, arXiv:1303.5076 
\bibitem[Fowler et al.(2007)]{Fow07} Fowler, J.~W., Niemack, M.~D., Dicker, S.~R., et al.\ 2007, Applied Optics, 46, 3444 
\bibitem[D{\"u}nner et al.(2013)]{Dunneretal2013} D{\"u}nner, R., Hasselfield, M., Marriage, T.~A., et al.\ 2013, \apj, 762, 10
\bibitem[Marriage et al.(2011)]{Marriageetal2011} Marriage, T.~A., Acquaviva, V., Ade, P.~A.~R., et al.\ 2011, \apj, 737, 61 
\bibitem[Menanteau et al.(2010)]{Menanteauetal2010} Menanteau, F., Gonz{\'a}lez, J., Juin, J.-B., et al.\ 2010, \apj, 723, 1523 
\bibitem[Menanteau et al.(2013)]{Menanteauetal2013} Menanteau, F., Sif{\'o}n, C., Barrientos, L.~F., et al.\ 2013, \apj, 765, 67 
\bibitem[Sif{\'o}n et al.(2013)]{Sifonetal2013} Sif{\'o}n, C., Menanteau, F., Hasselfield, M., et al.\ 2013, \apj, 772, 25 
\bibitem[Kirk et al.(2014)]{Kirketal2014} Kirk, B., Hilton, M., Cress, C., et al.\ 2014, arXiv:1410.7887 
\bibitem[Hilton et al.(2013)]{Hiltonetal2013} Hilton, M., Hasselfield, M., Sif{\'o}n, C., et al.\ 2013, \mnras, 435, 3469 
\bibitem[Sehgal et al.(2013)]{Sehgaletal2013} Sehgal, N., Addison, G., Battaglia, N., et al.\ 2013, \apj, 767, 38 
\bibitem[Hajian et al.(2011)]{HajCal} Hajian, A., Acquaviva, V., Ade, P.~A.~R., et al.\ 2011, \apj, 740, 86 
\bibitem[Dunkley et al.(2011)]{Dunkleyetal2011} Dunkley, J., Hlozek, R., Sievers, J., et al.\ 2011, \apj, 739, 52 
\bibitem[Hasselfield et al.(2013)]{Hasselfieldetal2013beam} Hasselfield, M., Moodley, K., Bond, J.~R., et al.\ 2013, \apjs, 209, 17
\bibitem[Marriage et al.(2011)]{Mar11} Marriage, T.~A., Baptiste Juin, J., Lin, Y.-T., et al.\ 2011, \apj, 731, 100 
\bibitem[Finkbeiner et al.(1999)]{Finkbeineretal1999} Finkbeiner, D.~P., Davis, M., \& Schlegel, D.~J.\ 1999, \apj, 524, 867 
\bibitem[Nozawa et al.(2006)]{Nozawaetal2006} Nozawa, S., Itoh, N., Suda, Y., \& Ohhata, Y.\ 2006, Nuovo Cimento B Serie, 121, 487 
\bibitem[Arnaud et al.(2005)]{Arnaudetal2005} Arnaud, M., Pointecouteau, E., \& Pratt, G.~W.\ 2005, \aap, 441, 893 
\bibitem[Bryan \& Norman(1998)]{Bryan-Norman1998} Bryan, G.~L., \& Norman, M.~L.\ 1998, \apj, 495, 80 
\bibitem[Duffy et al.(2008)]{Duffyetal2008} Duffy, A.~R., Schaye, J., Kay, S.~T., \& Dalla Vecchia, C.\ 2008, \mnras, 390, L64
\bibitem[Battaglia et al.(2010)]{Battagliaetal2010} Battaglia, N., Bond, J.~R., Pfrommer, C., Sievers, J.~L., \& Sijacki, D.\ 2010, \apj, 725, 91
\bibitem[Navarro et al.(1997)]{NFW1997} Navarro, J.~F., Frenk, C.~S., \& White, S.~D.~M.\ 1997, \apj, 490, 493
\bibitem[Arnaud et al.(2010)]{Arnaudetal2010} Arnaud, M., Pratt, G.~W., Piffaretti, R., et al.\ 2010, A \& A, 517, A92 
\bibitem[Sun et al.(2011)]{Sunetal2011} Sun, M., Sehgal, N., Voit, G.~M., et al.\ 2011, \apjl, 727, L49 
\bibitem[Planck Collaboration et al.(2013)]{Planck2013stack} Planck Collaboration, Ade, P.~A.~R., Aghanim, N., et al.\ 2013, \aap, 550, A131 
\bibitem[Planck Collaboration et al.(2013)]{Planck2012Coma} Planck Collaboration, Ade, P.~A.~R., Aghanim, N., et al.\ 2013, \aap, 554, A140 
\bibitem[Hajian et al.(2013)]{Hajianetal2013} Hajian, A., Battaglia, N., Spergel, D.~N., et al.\ 2013, \jcap, 11, 64 
\bibitem[Nagai et al.(2007)]{Nagaietal2007} Nagai, D., Kravtsov, A.~V., \& Vikhlinin, A.\ 2007, \apj, 668, 1 
\bibitem[Plagge et al.(2010)]{Plaggeetal2010} Plagge, T., Benson, B.~A., Ade, P.~A.~R., et al.\ 2010, \apj, 716, 1118 
\bibitem[Kaiser(1986)]{Kaiser1986} Kaiser, N.\ 1986, \mnras, 222, 323 
\bibitem[Voit(2005)]{Voit2005} Voit, G.~M.\ 2005, Reviews of Modern Physics, 77, 207 
\bibitem[Tinker et al.(2008)]{Tinkeretal2008} Tinker, J., Kravtsov, A.~V., Klypin, A., et al.\ 2008, \apj, 688, 709 
\bibitem[Molnar et al.(2009)]{Molnaretal2009} Molnar, S.~M., Hearn, N., Haiman, Z., et al.\ 2009, \apj, 696, 1640 
\bibitem[Sehgal et al.(2010)]{Sehgaletal2010} Sehgal, N., Bode, P., Das, S., et al.\ 2010, \apj, 709, 920 
\bibitem[Kesden et al.(2002)]{Kesdenetal2002} Kesden, M., Cooray, A., \& Kamionkowski, M.\ 2002, \prd, 66, 083007 
\bibitem[Trac et al.(2011)]{Tra11} Trac, H., Bode, P., \& Ostriker, J.~P.\ 2011, \apj, 727, 94 
\bibitem[Holder et al.(2007)]{Hol07} Holder, G.~P., McCarthy, I.~G., \& Babul, A.\ 2007, \mnras, 382, 1697 
\bibitem[Battaglia et al.(2012)]{Battagliaetal2012b} Battaglia, N., Bond, J.~R., Pfrommer, C., \& Sievers, J.~L.\ 2012, \apj, 758, 74 
\bibitem[Hu \& Kravtsov(2003)]{Hu-Kravtsov2003} Hu, W., \& Kravtsov, A.~V.\ 2003, \apj, 584, 702 
\bibitem[Takada \& Bridle(2007)]{Takada-Bridle2007} Takada, M., \& Bridle, S.\ 2007, \njp, 9, 446 
\bibitem[Takada \& Spergel(2013)]{Takada-Spergel2013} Takada, M., \& Spergel, D.~N.\ 2013, arXiv:1307.4399 
\bibitem[Zhang \& Sheth(2007)]{Zhang-Sheth2007} Zhang, P., \& Sheth, R.~K.\ 2007, \apj, 671, 14 
\bibitem[Limber(1954)]{Limber1954} Limber, D.~N.\ 1954, \apj, 119, 655
\bibitem[Tinker et al.(2010)]{Tinkeretal2010} Tinker, J.~L., Robertson, B.~E., Kravtsov, A.~V., et al.\ 2010, \apj, 724, 878 
\bibitem[Eifler et al.(2009)]{Eifleretal2009} Eifler, T., Schneider, P., \& Hartlap, J.\ 2009, \aap, 502, 721 
\bibitem[Shaw et al.(2009)]{Shawetal2009} Shaw, L.~D., Zahn, O., Holder, G.~P., \& Dor{\'e}, O.\ 2009, \apj, 702, 368 
\bibitem[Das et al.(2011)]{Dasetal2011b} Das, S., Marriage, T.~A., Ade, P.~A.~R., et al.\ 2011, \apj, 729, 62 
\bibitem[Das et al.(2011)]{Dasetal2011} Das, S., Sherwin, B.~D., Aguirre, P., et al.\ 2011, Physical Review Letters, 107, 021301 
\bibitem[Huffenberger \& Seljak(2005)]{Huffenberger-Seljak2005} Huffenberger, K.~M., \& Seljak, U.\ 2005, \na, 10, 491 
\bibitem[Sayers et al.(2013)]{Sayersetal2013} Sayers, J., Mroczkowski, T., Czakon, N.~G., et al.\ 2013, \apj, 764, 152 
\bibitem[Diego \& Partridge(2010)]{Diego-Partridge2010} Diego, J.~M., \& Partridge, B.\ 2010, \mnras, 402, 1179 
\bibitem[Best et al.(2007)]{Bestetal2007} Best, P.~N., von der Linden, A., Kauffmann, G., Heckman, T.~M., \& Kaiser, C.~R.\ 2007, \mnras, 379, 894 
\bibitem[Gralla et al.(2011)]{Grallaetal2011} Gralla, M.~B., Gladders, M.~D., Yee, H.~K.~C., \& Barrientos, L.~F.\ 2011, \apj, 734, 103 
\bibitem[Lin et al.(2009)]{Linetal2009} Lin, Y.-T., Partridge, B., Pober, J.~C., et al.\ 2009, \apj, 694, 992 
\bibitem[George et al.(2014)]{Georgeetal2014} George, E.~M., Reichardt, C.~L., Aird, K.~A., et al.\ 2014, arXiv:1408.3161 
\bibitem[Lindner et al.(2014)]{Lindneretal2014} Lindner, R., et al. \emph{submitted}
\bibitem[Liu et al.(2014)]{Liuetal2014} Liu, J., Mohr, J., Saro, A., et al.\ 2014, arXiv:1407.7520 
\bibitem[Bleem et al.(2013)]{Bleem2013} Bleem L. E., 2013, PhD thesis, The University of Chicago
\bibitem[Righi et al.(2008)]{Righietal2008} Righi, M., Hern{\'a}ndez-Monteagudo, C., \& Sunyaev, R.~A.\ 2008, \aap, 478, 685 
\bibitem[Vikhlinin et al.(2009)]{Vikhlininetal2009} Vikhlinin, A., Kravtsov, A.~V., Burenin, R.~A., et al.\ 2009, \apj, 692, 1060 
\bibitem[Spergel et al.(2013)]{Spergeletal2013} Spergel, D., Flauger, R., \& Hlozek, R.\ 2013, arXiv:1312.3313
\bibitem[Planck Collaboration et al.(2014)]{Planck2013likelihood} Planck Collaboration, Ade, P.~A.~R., Aghanim, N., et al.\ 2014, \aap, 571, AA15 




\end{thebibliography}
\end{document}